\documentclass[%
 aip,
 amsmath,amssymb,
 reprint,%
]{revtex4-2}

\usepackage{graphicx}
\usepackage{dcolumn}
\usepackage{bm}

\usepackage[utf8]{inputenc}
\usepackage[T1]{fontenc}
\usepackage{mathptmx}
\usepackage{etoolbox}
\usepackage{color}

\makeatletter
\def\@email#1#2{%
 \endgroup
 \patchcmd{\titleblock@produce}
  {\frontmatter@RRAPformat}
  {\frontmatter@RRAPformat{\produce@RRAP{*#1\href{mailto:#2}{#2}}}\frontmatter@RRAPformat}
  {}{}
}%
\makeatother
\begin{document}

\preprint{AIP/123-QED}

\title{G\ae{}nice: a general model for magnon band structure of artificial spin ices}
\author{Ghanem Alatteili}
\author{Victoria Martinez}
\author{Alison Roxburgh}
\affiliation{Center for Magnetism and Magnetic Nanostructures, University of Colorado Colorado Springs, Colorado Springs, CO 80918, USA}

\author{Jack C. Gartside}
\affiliation{Blackett Laboratory, Imperial College London, Prince Consort Road, London SW7 2AZ, UK}

\author{Olle G. Heinonen}
 \altaffiliation[Current and permanent address: ]{Seagate Technology, 7801 Computer Ave., Bloomington, MN 55435}
\affiliation{Materials Science Division, Argonne National Laboratory, Lemont, Illinois 60439, USA}

\author{Sebastian Gliga}
\affiliation{Swiss Light Source, Paul Scherrer Institute, 5232 Villigen PSI, Switzerland}

\author{Ezio Iacocca}
 \email{eiacocca@uccs.edu}
\affiliation{Center for Magnetism and Magnetic Nanostructures, University of Colorado Colorado Springs, Colorado Springs, CO 80918, USA}

\date{\today}

\begin{abstract}
Arrays of artificial spin ices exhibit reconfigurable ferromagnetic resonance frequencies that can be leveraged and designed for potential applications. However, analytical and numerical studies of the frequency response of artificial spin ices have remained somewhat limited due to the need of take into account nonlocal dipole fields in theoretical calculations or by long computation times in micromagnetic simulations. Here, we introduce G\ae{}nice, a framework to compute magnon dispersion relations of arbitrary artificial spin ice configurations. G\ae{}nice makes use of a tight-binding approach to compute the magnon bands. It also provides the user complete control of the interaction terms included, e.g., external field, anisotropy, exchange, and dipole, making it useful also to compute ferromagnetic resonances for a variety of structures, such as multilayers and ensembles of weakly or non-interacting nanoparticles. Because it relies on a semi-analytical model, G\ae{}nice is computationally inexpensive and efficient, making it an attractive tool for the exploration of large parameter spaces.
\end{abstract}

\maketitle

%
\section{\label{sec:intro}Introduction\protect\\}
Artificial spin ices (ASIs) are systems of structured nanomagnets arranged in periodic patterns that are magneto-statically coupled. ASIs were originally designed  
to mimic the behavior of natural spin ice materials~\cite{Skaervo2019} ,in order to explore the fundamental principles of frustrated magnetism. Frustration arises from competing magnetic interactions that cannot all be simultaneously minimized~\cite{Heyderman2013}, leading to highly degenerate states. ASIs can be also considered as magnonic crystals~\cite{Gliga2013,Gliga2020,Lendinez2019} exhibiting reconfigurable magnonic modes~\cite{Gartside2018,Mamica2018,Arroo2019,Dion2019,Lendinez2019,Iacocca2020,MIcaletti2023}, nonlinear scattering~\cite{Lendinez2023}, band structure~\cite{Iacocca2016,Lasnier2020,Montoncello2023}, and hybrid modes~\cite{Graczyk2018,Negrello2022}.


The arrangement of magnetic elements in a square lattice, known as square ice~\cite{Wang2006},has been a test bed for the investigation of ASIs as magnonic crystals because its relative simplicity allows for the understanding of the fundamental physical phenomena. Analytically, square ices have proven  promising for reconfigurable magnonics because of the mode-dependent magnon modes predicted~\cite{Iacocca2016,Lasnier2020} as well as evidence of topological modes~\cite{Iacocca2017c}. However, the study of 
similar effects in other geometries remains limited to date. Experimentally, this is partly because of the large number of geometries to explore~\cite{Skaervo2019} and the technical challenges to investigate wavevector- or spatially-resolved magnons in a nanopatterned structure by, e.g. by Brillouin light scattering~\cite{Dion2023}. From a numerical point of view, simulations  using micromagnetic modeling~\cite{Abert2019} are very time-consuming and often require large memory allocations to investigate long-wavelength magnons that are easily excited by microwave antennas in experiments. 
This means that more exotic geometries, both in 2D~\cite{Skaervo2019} and 3D~\cite{May2019,May2021,Sahoo2021}, have been slower to materialize due to the lack of an efficient predictive tool for magnetization dynamics.

Here, we present G\ae{}nice, 
a general formalism to compute the magnon dispersion relation for arbitrary ASI geometries. 
The formalism is based on a Holstein-Primakoff transformation~\cite{Slavin2009} to obtain an eigenvalue problem that can be solved numerically with little computational cost~\cite{Iacocca2016}. The main difference between G\ae{}nice and other analytical methods is its generalization to arbitrary nanomagnet orientations and magnetization states with an automatic determination of the first Brillouin zone (FBZ). 
We expect that G\ae{}nice can serve as a numerically efficient and computationally accurate tool to predict magnonic functionality for ASIs and to direct more detailed studies of promising geometries using micromagnetic simulations and experiments. In other words, we envision G\ae{}nice as a tool to quickly explore the parameter space of distinct ASI geometries and identify potentially interesting regimes that can be then further explored with traditional computational and experimental methods~\cite{Grimsditch2004,Neusser2011,Rychly2015,Gubbiotti2018,Lisiecki2019}

The remainder of the paper is organized as follows. In section II, we describe the general formulation of the problem. The energy terms considered and their implementation are detailed in section III. In section IV, we demonstrate the functionality of G\ae{}nice by computing simple Kittel modes and ferromagnetic resonance (FMR) modes in a variety of linear arrays of nanomagnets and the two fundamental ASI configurations: square ice and Kagome ice.

\section{\label{sec:level1}Generalized analytical model}



We begin our description from the conservative Larmor torque equation
\begin{equation}
\label{eq:1}
    \frac{\partial \mathbf{m}}{\partial t} = -\gamma\mu_0\mathbf{m}\times\mathbf{H}_\mathrm{eff},
\end{equation}
where $\mathbf{m}$ is the normalized magnetization vector with $|\mathbf{m}|=1$, $\gamma$ is the gyromagnetic ratio, and $\mu_0$ is the vacuum permeability. The effective field $\mathbf{H}_\mathrm{eff}$ contains physical terms and phenomena relevant to the magnetic material and interfaces which are described within the context of our eigenvalue solver in section~\ref{sec3}. Note that we neglect damping here given that we are interested in resonant, propagating modes.

\begin{figure}[t]
\centering \includegraphics[width=2in]{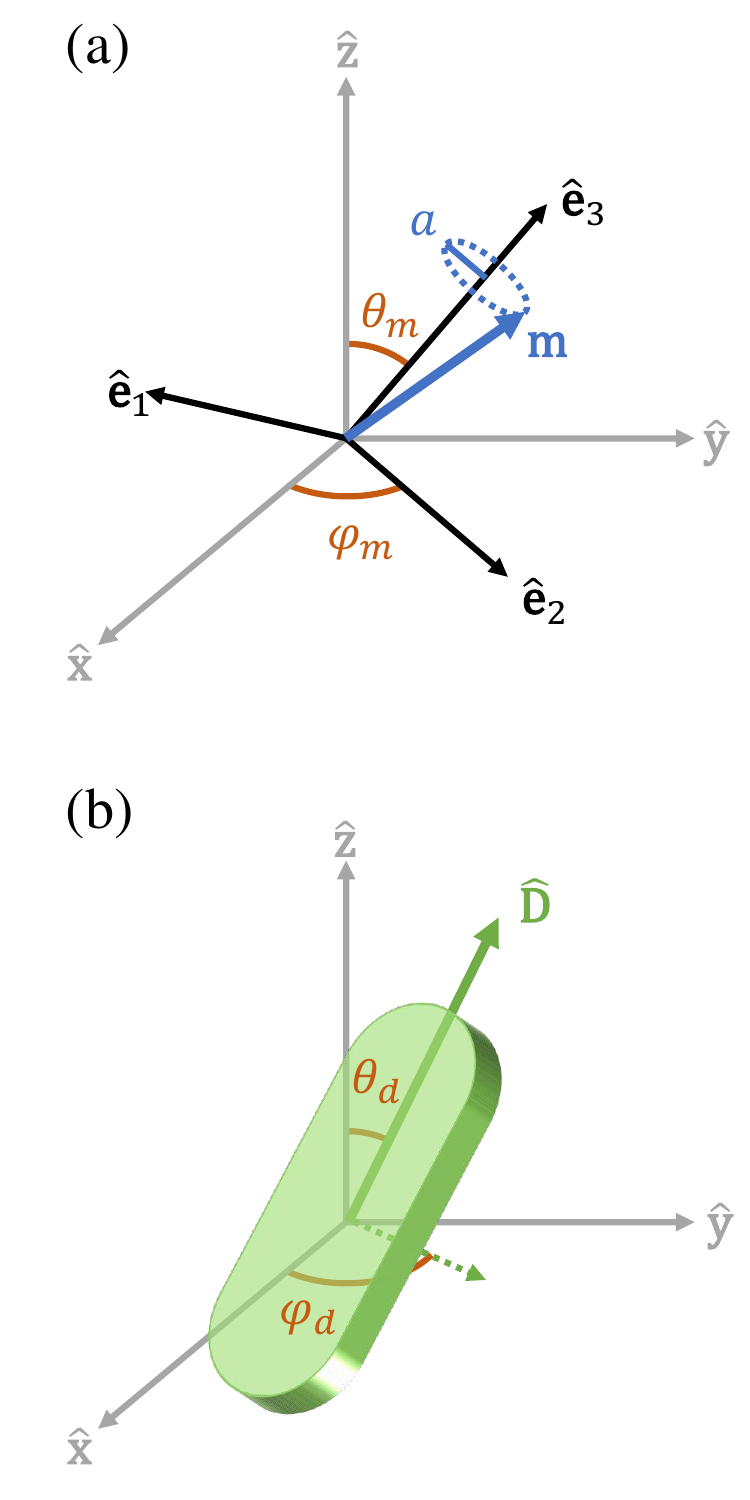}
\caption{ \label{fig:coords} (a) The magnetization vector rotated by the polar and azimuthal angles $(\theta_m,\varphi_m)$. The rotated frame $(\hat{\mathbf{e}}_1,\hat{\mathbf{e}}_2,\hat{\mathbf{e}}_3)$ is defined relative to the equilibrium orientation of the magnetization vector $\mathbf{m}$ (b) Nanomagnet orientation relative to the Cartesian coordinate system. The unit vector $\hat{\mathbf{D}}$ indicates the direction of the nanomagnet's long axis and is defined by the angles $\theta_d$ and $\varphi_d$.} 
\end{figure}

For small-amplitude excitations, such as magnons, the Larmor torque equation can be rewritten as a Hamiltonian set of equations using a Holstein-Primakoff transformation of the \emph{complex} small amplitude $a$~\cite{Slavin2009}
\begin{equation}
\label{eq:2}
    a = \frac{m_1 + im_2}{\sqrt{2(1+m_3^2)}},
\end{equation}
where $\mathbf{m}=m_1\hat{\mathbf{e}}_1+m_2\hat{\mathbf{e}}_2+m_3\hat{\mathbf{e}}_3$ is the magnetization vector expressed in a coordinate system where $\hat{\mathbf{e}}_3$ 
defines the equilibrium orientation of the magnetization vector 
and $\hat{\mathbf{e}}_1\times\hat{\mathbf{e}}_2=\hat{\mathbf{e}}_3$. An illustration of this basis is shown in Fig.~\ref{fig:coords}(a).

From Eq.~\ref{eq:2}, we can relate the complex amplitudes to the magnetization vector in the $\hat{\mathbf{e}}$ basis as
\begin{subequations}
\begin{eqnarray}
\label{eq:2_1}
    m_1 &=& \sqrt{1-|a|^2}(a+a^*)\approx\left(1-\frac{|a|^2}{2}\right)(a+a^*)\\
\label{eq:2_2}
    m_2 &=& i\sqrt{1-|a|^2}(a-a^*)\approx i\left(1-\frac{|a|^2}{2}\right)(a-a^*)\\
\label{eq:2_3}
    m_3 &=& (1-2|a|^2)
\end{eqnarray}
\end{subequations}

Using the transformation of Eq.~\ref{eq:2}, we can approximately rewrite Eq.~\ref{eq:1} as a Hamiltonian system for the complex amplitude $a$
\begin{equation}
\label{eq:3}
    \frac{\partial a}{\partial t} = -i \frac{\partial}{\partial a^*} a \mathcal{H} a^\dagger
\end{equation}
and the Hamiltonian is defined over a magnetic volume as
\begin{equation}
\label{eq:4}
    \mathcal{H} = -\mu_0M_s\int \mathbf{H}(\mathbf{m})\cdot{\mathbf{m}}dA,
\end{equation}

To describe the magnon band structure of an ensemble of nanomagnets in an ASI, the auto-oscillator model can be generalized to an array of complex amplitudes, as shown in Ref.~\onlinecite{Iacocca2016}. To account for the bending of the magnetization at edges of the nanomagnets~\cite{Gliga2015}, we divide each nanomagnet into 3 \emph{macrospins}. This is an important assumption in our model, making it valid for magnetic elements with sizes of the order of hundreds of nanometers. Therefore, given $N$ nanomagnets in the unit cell of the ASI, 
we define the complex amplitude array $\underline{a}=[a_1~a_2~ ...~ a_{3N}]$ and the $2(3N)\times 2(3N)$ Hamiltonian matrix $\underline{\mathcal{H}}$ so that the generalized {Hamiltonian} becomes
\begin{equation}
\label{eq:5}
    \frac{d}{dt}\underline{a} = -i\frac{d}{d\underline{a}^*}\begin{bmatrix}\underline{a}&\underline{a}^*\end{bmatrix}\underline{\mathcal{H}}\begin{bmatrix}\underline{a}\\\underline{a}^*\end{bmatrix},
\end{equation}

The Hamiltonian matrix is further divided as
\begin{equation}
\label{eq:5_6}
    \mathcal{H} = \begin{bmatrix}\mathcal{H}^{(1,2)}&\mathcal{H}^{(2,2)}\\\mathcal{H}^{(1,1)}&\mathcal{H}^{(2,1)}\end{bmatrix},
\end{equation}
where $\mathcal{H}^{(1,1)}=(\mathcal{H}^{(2,2)})^*$ and $\mathcal{H}^{(1,2)}=(\mathcal{H}^{(2,1)})^*$ by symmetry of the Hamiltonian equations. As further discussed below, this system describes bosonic excitation (magnons) so that the eigenvalue problem can be solved using Colpa's grand dynamical matrix that ensures complex conjugate eigenvalues~\cite{Colpa1978}.

\subsection{\label{sec:coord}Coordinate system}

Our 
framework relies on a Cartesian coordinate system where the polar angle $\theta=0$ and the azimuth angle $\varphi=0$ define the $z$-axis, as shown in Fig.~\ref{fig:coords}(a). Once this coordinate system is established, applying Eq.~\ref{eq:1} requires a rotation to the coordinate system defined by $(\hat{\mathbf{e}}_1,\hat{\mathbf{e}}_2,\hat{\mathbf{e}}_3)$, where the direction of $\hat{\mathbf{e}}_3$ is parallel to the equilibrium orientation of the magnetization vector at any given point in space.

This implies that coordinate transformations must be performed locally for both the magnetization vector and the effective field for an arbitrary array of magnetization vectors. We define the local rotation matrix
\begin{equation}
\label{eq:8}
    R(\theta,\varphi)=\begin{bmatrix}\cos{\varphi}\cos\theta&\sin\varphi\cos\theta&-\sin\theta\\-\sin\varphi&\cos\varphi&0\\ \cos\varphi\sin\theta&\sin\varphi\sin\theta&\cos\theta\end{bmatrix}.
\end{equation}
It is important to note that $R^{-1}(\theta,\varphi)=R(\theta,\varphi)$.

{An} arbitrary orientation of nanomagnets is considered, defined by a unit vector $\hat{\mathbf{D}}$. The direction of $\hat{\mathbf{D}}$ is parameterized by the polar and azimuth angles $\theta_d$ and $\varphi_d$ as shown in Fig.~\ref{fig:coords}(b). When $\theta_d=0$ and $\varphi_d=0$, the nanomagnet is aligned along the $z$-axis, and its thickness is aligned along the $x$-axis. In this case, the angles $\theta_d$ and $\varphi_d$ represent pitch and yaw, respectively.

\subsection{\label{sec:level3}Eigenvalue problem}
The magnon dispersion relation $\omega(k)$ is 
obtained from Eq.~\ref{eq:5} by invoking Bloch's theorem $a\rightarrow ae^{i\omega}$ and 
Colpa's grand dynamical matrix~\cite{Colpa1978}
\begin{equation}
\label{eq:6}
    \omega\underline{\Psi} \propto \begin{bmatrix}\mathcal{H}^{(1,2)}&-(\mathcal{H}^{(1,1)})^*\\\mathcal{H}^{(1,1)}&-(\mathcal{H}^{(1,2)})^*\end{bmatrix}\underline{\Psi},
\end{equation}
where $\underline{\Psi}$ is an array of eigenvectors. 
By introducing the proportionality factor $\gamma/(2VM_s)$, where $V$ is a volume and $M_s$ is the saturation magnetization, Eq.~\ref{eq:6} becomes an equality. The volume and saturation magnetization will be associated with a magnetization vector to allow for maximal flexibility of the model. Therefore, we will use the convention $\Omega = \gamma/(2VM_s)\mathcal{H}$ to obtain
\begin{equation}
\label{eq:7}
    \omega\underline{\Psi} = \begin{bmatrix}\underline{\Omega}^{(1,2)}&-(\underline{\Omega}^{(1,1)})^*\\\underline{\Omega}^{(1,1)}&-(\underline{\Omega}^{(1,2)})^*\end{bmatrix}\underline{\Psi},
\end{equation}

The eigenvalue problem of Eq.~\ref{eq:7} can be solved numerically by standard methods. This formulation has been implemented for square ice
~\cite{Iacocca2016,Iacocca2017c} and it has been referred to as semi-analytical because 
the Hamiltonian matrix $\mathcal{H}$ is derived analytically and only the eigenvalue problem is solved numerically. Here, we derive the Hamiltonian matrices for arbitrary ASI configurations, so that the matrix is built in an automated way for any number of nanomagnets in a unit cell.

\section{\label{sec3}Effective field}

The effective field is G\ae{}nice's 
core, which gives rise to the Hamiltonian matrix. In essence, the effective field is divided into two groups of physical effects; local and non-local
\begin{equation}
\label{eq:9}
    \mathbf{H}_\mathrm{eff} = \mathbf{H}_\mathrm{l} + \mathbf{H}_\mathrm{nl}.
\end{equation}

In its current implementation, G\ae{}nice includes a uniform external magnetic field and anisotropy field as local contributions; and exchange interaction and dipole-dipole interaction as non-local fields from a point of view that the macrospins of the nanomagnets are coupled. In other words, these fields lead to finite non-diagonal elements in the Hamiltonian block matrices. The dipole-dipole contribution is fundamental to ASIs and is detailed below. In the same manner, it is possible to extend G\ae{}nice to include other field contributions, such as magnetocrystalline anisotropy, Dzyaloshinskii-Moriya interaction, and RKKY exchange.

In the following subsections, we express these field contributions in the form given in Eq.~\ref{eq:7}. All fields are defined in the Cartesian coordinate system and rotated to the local magnetization coordinates described in section~\ref{sec:coord}.

\subsection{External Field}
A uniform external field, $\mathbf{H}_0$, leads to the energy
\begin{equation}
\label{eq:10}
    E_\mathrm{0} = -V\mu_0M_s(R(\theta_m,\varphi_m)\cdot\mathbf{H}_0)^T\cdot\mathbf{m}.
\end{equation}

The only quadratic term in $a$ in Eq.~\ref{eq:10} is parallel to $\hat{\mathbf{e}}_3$. Therefore, we can express the frequency contribution due to an external field as
\begin{equation}
\label{eq:11}
    \Omega_0 = \gamma\mu_0|a|^2(R(\theta_m,\varphi_m)\cdot\mathbf{H}_0)^T\cdot\hat{\mathbf{e}}_3,
\end{equation}
with diagonal and off-diagonal blocks
\begin{subequations}
\begin{eqnarray}
\label{eq:21_1}
    \underline{\Omega}_0^{(1,1)}  &=& 0\\
\label{eq:21_2}
    \underline{\Omega}_0^{(1,2)}  &=& \frac{\gamma\mu_0}{2}\left[(R(\theta_m,\varphi_m)\cdot\mathbf{H}_0)^T \cdot\hat{\mathbf{e}}_3\right]\mathbf{I},
\end{eqnarray}
\end{subequations}
where $\mathbf{I}$ is the identity matrix.

\subsection{Demagnetization Field}

The demagnetization (demag) field is determined from 
the shape of the magnetic element, which is an accurate approximation for soft magnets, such as Permalloy. 
We consider {a} demag tensor $\underline{D}$ that we approximate with diagonal demagnetizing factors $D_1<D_2<D_3$  such that
\begin{equation}
\label{eq:12}
    \underline{D} = \begin{bmatrix}D_3&0&0\\0&D_2&0\\0&0&D_1\end{bmatrix}.
\end{equation}
This approximation is consistent with the notion of macrospin elements, i.e., similar to the general ellipsoid~\cite{Osborn1945}.

This definition follows from the nanomagnet's orientation in the Cartesian coordinate system whereby the easy axis lies along the $z$-axis and the hard axis along the $x$-axis. The demagnetizing factors can be found in a variety of ways. Analytical expressions are available for oblate nanomagnets~\cite{Osborn1945} and for rectangular prisms~\cite{Aharoni1998}. Demag factors can also be obtained by fitting FMR from simulated nanomagnets and empirical expressions can be found for a range of aspect ratios~\cite{Martinez2023} G\ae{}nice currently supports the oblate nanomagnet analytical expressions given by.
\begin{subequations}
\begin{eqnarray}
\label{eq:13_1}
    D_1 &=& \frac{t\sqrt{1-e^2}(K-E)}{l e^2},\\
\label{eq:13_2}
    D_2 &=& \frac{t\left(E-(1-e^2)K\right)}{l e^2\sqrt{1-e^2}},\\
\label{eq:13_3}
    D_3 &=& \frac{1-t E}{l\sqrt{1-e^2}},
\end{eqnarray}
\end{subequations}
where $K$ and $E$ are the complete elliptic integrals of the first and second kind and $e=\sqrt{1-(w/l)^2}$. In the limit of a circular nanomagnet, one must consider $D_1=D_2=0$ and $D_3=1$ to avoid a numerical singularity.

The anisotropy energy is expressed as 
\begin{equation}
\label{eq:14}
E_\mathrm{an} = V\mu_0M_s^2\vec{m}\cdot\underline{D}\cdot\vec{m}^T,
\end{equation}
  
We rotate the demagnetization tensor using Eq.~\ref{eq:8} and the director vector $\hat{\mathbf{D}}$ to align it with the magnetization direction in the Cartesian reference frame 
\begin{equation}
\label{eq:15}
    \underline{C} = R(\theta_m-\theta_d,\varphi_m-\varphi_d)\cdot\underline{D}\cdot R(\theta_m-\theta_d,\varphi_m-\varphi_d)^{-1},
\end{equation}   
resulting in the nanomagnet-dependent Hamiltonian matrix
 \begin{equation}
 \label{eq:16}
    \mathbf{\mathcal{H}}_\mathrm{an}
    = VM_s^2 \vec{m}\cdot\underline{C}\cdot\vec{m}^T.
\end{equation}

We note that this matrix is a $3\times3$ block that is defined for each nanomagnet. Expressing Eq.~\ref{eq:16} as a function of the complex amplitudes $a$, ultimately results in the diagonal Hamiltonian block matrices
\begin{subequations}
\begin{eqnarray}
    \label{eq:28_1}
\underline{\Omega}_{an}^{(1,1)} &=& \frac{\gamma\mu_0M_s}{2}\left[C_{11}-C_{22}+ i(C_{12}+C_{21})\right]\mathbf{I},\\
     \label{eq:28_2}
\underline{\Omega}_{an}^{(1,2)} &=& \frac{\gamma\mu_0M_s}{2}\left[C_{11}-C_{22} - 2C_{33}\right]\mathbf{I},
     \end{eqnarray}
    \end{subequations}
where the factors $C_{ij}$ are the coefficients of the matrix $\underline{C}$.

\subsection{Exchange}

\begin{figure}[t]
\centering \includegraphics[width=3in]{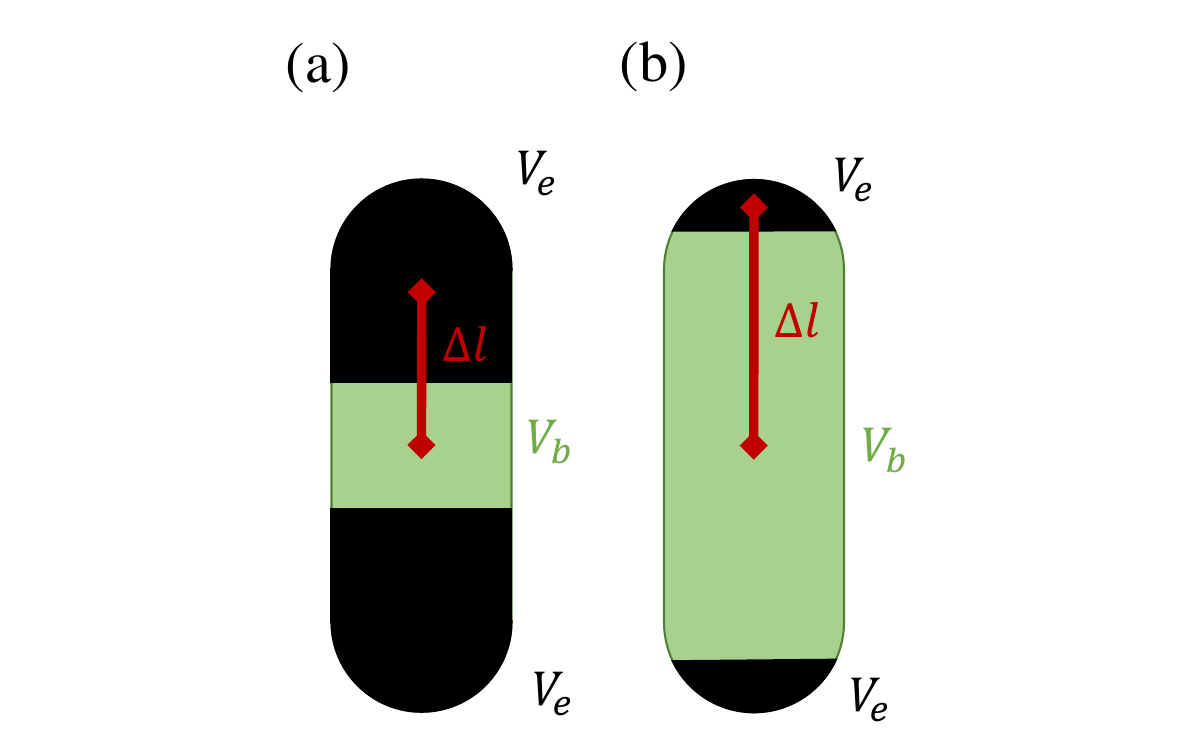}
\caption{ \label{fig:Macrospins} Symmetric splitting of stadium-shaped nanomagnets, where we discern between a large edge volume (a) and a small edge volume (b), relative to the semi-circular edges. The bulk and edge volumes, $V_b$ and $V_e$, respectively, are uniquely determined by the parameter $\Delta l$. }
\end{figure}

We include exchange interaction as a minimal model for edge bending in the magnetization of tightly packed nanomagnets~\cite{Gliga2015,Iacocca2016}. The nanomagnet is split into three regions, and we use 
an effective exchange energy to parameterize the exchange interaction. The nanomagnet splitting is shown in Fig.~\ref{fig:Macrospins} for the cases where the edge modes are (a) larger or (b) smaller than the stadium's semi-circular edges. It is assumed that the edge modes are symmetric in volume. We refer to Appendix~\ref{app:exchange} for details. Here, we report the final form of the block matrices used in the eigenvalue problem.

We consider that the nanomagnet is split in a bulk macrospin with volume $V_b$ and two edge macrospins with identical volumes $V_e$, satisfying $V=V_b+2V_e$. The volumes are uniquely determined by the parameter $\Delta l$ defined as the length from the geometric center of the nanomagnet to the center of the edge volume. The default value $\Delta l=(2l-w)/4$ is defined when the edge volume is exactly contained at the semi-circular edges of stadium-shaped nanomagnets. However, this parameter can be tuned.

The exchange energy is therefore defined as a pair-wise interaction between a bulk macrospin $b$ and an edge macrospin $e+$ (upper) and $e-$ (lower)
\begin{equation}
\label{eq:23}
    E_\mathrm{ex}^{(b,e)}=-\frac{J}{2}\mathbf{m}_b^T\cdot R(\theta_{m_e}-\theta_{m_b},\varphi_{m_e}-\varphi_{m_b})\cdot \mathbf{m}_e,
\end{equation}
where the exchange factor $J$ is given by
\begin{equation}
\label{eq:21}
    J = \frac{2}{\Delta l^2}\left(V_e+\frac{V_b}{2}\right).
\end{equation}

Recognizing that the exchange interaction only occurs for neighboring macrospins within a nanomagnet, we define the $3\times3$ exchange energy blocks per nanomagnet $N$
\begin{subequations}
\label{eq:26}
\begin{eqnarray}
\label{eq:26_1}
    \underline{E}_{ex,N}^{(1,1)} &=& -\frac{J}{2}\begin{bmatrix}0 & E_1^{(b,e^+)}/V_e & 0\\E_1^{(e^+,b)}/V_b&0&E_1^{(b,e^-)}/V_b\\0 & E_1^{(e^-,b)}/V_e&0\end{bmatrix},\\
\label{eq:26_2}
    \underline{E}_{ex,N}^{(1,2)} &=& -\frac{J}{2}\begin{bmatrix}-\Sigma E^+ & E_2^{(b,e^+)}/V_e & 0\\E_2^{(e^+,b)}/V_b&-\Sigma E^+-\Sigma E^-&E_2^{(b,e^-)}/V_b\\0 & E_2^{(e^-,b)}/V_b&-\Sigma E^-\end{bmatrix},
\end{eqnarray}
\end{subequations}
where we define $\Sigma E^e=E_3^{b,e}/V_e+E_3^{e,b}/V_b$, and
\begin{subequations}
\begin{eqnarray}
\label{eq:25_1}
    E_1^{b,e} &=& R^{1,1}-R^{2,2}+i(R_{1,2}+R^{2,1}),\\
\label{eq:25_2}
    E_2^{b,e} &=& R^{1,1}+R^{2,2}-i(R_{1,2}-R^{2,1}),\\
\label{eq:25_3}
    E_3^{b,e} &=& R^{3,3},
\end{eqnarray}
\end{subequations}

With these blocks, the frequency contribution due to exchange is written with the block-diagonal matrices as
\begin{subequations}
\begin{eqnarray}
\label{eq:27_1}
    \underline{\Omega}_{ex}^{(1,1)} &=& \frac{\gamma}{2M_s}\begin{bmatrix}\underline{E}_{ex,1}^{(1,1)} & & \\
    & \ddots & \\
    & & \underline{E}_{ex,N}^{(1,1)}\end{bmatrix}\\
\label{eq:27_2}
    \underline{\Omega}_{ex}^{(1,2)} &=& \frac{\gamma}{2M_s}\begin{bmatrix}\underline{E}_{ex,1}^{(1,2)} & & \\
    & \ddots & \\
    & & \underline{E}_{ex,N}^{(1,2)}\end{bmatrix}
\end{eqnarray}
\end{subequations}

\subsection{Dipole field}

The dipole field is essential to compute the spin-wave band structure for ASIs. 
We distinguish two contributions to the dipole field: a static contribution originating from the equilibrium magnetization, and a dynamic contribution originating from the long-range dynamics of macrospins. An analogous way to phrase this, is that we consider a perturbation to the dynamical matrix where the zeroth order term is the static stray field from the magnetization and the first-order correction is the dipole-dipole contribution.

\subsubsection{Static contribution}

To compute the static contribution of the dipole, we consider the stray field from each nanomagnet in the ASI on a macrospin $i$. As an approximation, we implemented the analytical expressions of the stray field from a rectangular prism derived by R. Engel-Herbert and T. Hesjedal~\cite{Engel2005}. The resulting field due to nanomagnet $n$ is $\mathbf{H}_{stray,N}^{(i)}$ and is computed as a function of the center position of the nanomagnet $n$ and the position of the \emph{macrospin} $i$. The analytical expressions derived in Ref.~\cite{Engel2005} are written in Appendix~\ref{sec:appstray}. Essentially, this computation provides a local field source for macrospin $i$ so that it contributes to the Hamiltonian matrix as an external magnetic field.

A subtle difference between a truly external field and the stray field is that we need to scale the latter to the fractional volume of the macrospin it is acting upon. In other words, we impose that the total energy on the target nanomagnet due to nanomagnet $n$ is conserved
\begin{equation}
    \label{eq:127}
    E = \mu_0M_s\frac{\sum_{i=1}^3V_i\mathbf{H}_{stray,n}^{(i)}\cdot\mathbf{m}_i}{V}.
\end{equation}

As a consequence, the contribution to the matrix becomes
\begin{subequations}
\begin{eqnarray}
\label{eq:121_1}
    \underline{\Omega}_{stray}^{(1,1)}  &=& 0,\\
\label{eq:121_2}
    \underline{\Omega}_{stray}^{(1,2)}  &=& \frac{\gamma\mu_0}{2}\sum_{\tau_1,\tau_2}^\mathrm{ASI}\left[\sum_{i=1}^3\frac{V_i}{V}R(\theta_m,\varphi_m)\cdot\mathbf{H}_{stray,n}^{(i)})^T \cdot\hat{\mathbf{e}}_3\right]\mathbf{I},
\end{eqnarray}
\end{subequations}
where $\tau_1$ and $\tau_2$ are integer numbers so that $\tau_1=\tau_2=0$ is the unit cell. The maximum value of $\tau_1$ and $\tau_2$ is capped so that the long-range dipole contributions converge with sufficient numerical accuracy. G\ae{}nice permits to specify either the maximum value for $\tau_1$ and $\tau_2$ or to expand the lattice until numerical accuracy is achieved.

\subsubsection{Dynamic contribution}

The dipole field of macrospin $j$ acting on macrospin $i$ 
is calculated using the following expression. 
\begin{equation}
\label{eq:28}
    \mathbf{H}_{\mathrm{d},ij} = \frac{V_jM_{s,j}}{4\pi}\left[\frac{3\mathbf{r}_{i,j}(\mathbf{r}_{i,j}\cdot\mathbf{m}_j)}{|\mathbf{r}_{i,j}|^5}-\frac{\mathbf{m}_j}{|\mathbf{r}_{i,j}|^3}\right],
\end{equation}
where, $\mathbf{r}_{i,j}$ is the distance between the two macrospins $i$ and $j$.

We adopt a tight-binding-like approach for periodic structures whereby the dipole field is computed within and between unit cells. Therefore, the long-range terms collapse into a single Hamiltonian matrix. Identifying each macrospin's spatial position within a nanomagnet is critical. Hence, we adopt the convention that the bulk macrospin is located at the nanomagnet's geometric center, denoted by $\mathbf{X}_b$. The edge macrospins can be 
computed by
\begin{equation}
\label{eq:29_pre}
    \mathbf{X}_{e^\pm} = \mathbf{X}_b \pm \Delta l \hat{\mathbf{D}}.
\end{equation}

 We consider the periodic structure established by the 
 translation vectors $a_1$ and $a_2$. 
 In general, the distance between macrospins $i$ and $j$ 
 is given by
\begin{equation}
\label{eq:29}
    \mathbf{r}_{ij}^{(\tau_1,\tau_2)} = \mathbf{X}_i-\mathbf{X}_j-(\tau_1\mathbf{a}_1+\tau_2\mathbf{a}_2).
\end{equation}

Therefore, the total nonlocal dipole field acting on a macrospin $i$ can be written as
\begin{eqnarray}
\label{eq:30}
    \mathbf{H}_{\mathrm{d},ij} &=& \frac{1}{4\pi}\sum_{\tau_1,\tau_2}^\mathrm{ASI}\sum_j^\mathrm{U.C.}V_jM_{s,j}\Big[\frac{3\mathbf{r}_{ij}^{(\tau_1,\tau_2)}(\mathbf{r}_{ij}^{(\tau_1,\tau_2)}\cdot\mathbf{m}_j^{(\tau_1,\tau_2)})}{|\mathbf{r}_{ij}^{(\tau_1,\tau_2)}|^5}\nonumber\\&-&\frac{\mathbf{m}_j}{|\mathbf{r}_{ij}^{(\tau_1,\tau_2)}|^3}\Big],
\end{eqnarray}


Because the distances in Eq.~\eqref{eq:29} are computed in the natural Cartesian coordinates, we need to rotate {into} the basis of each macrospin to compute the products as a function of coupled complex amplitudes. For this we define two distances in the rotated reference frame

\begin{subequations}
\begin{eqnarray}
\label{eq:31_1}
    \mathbf{\rho}_{ij} &=& R(\theta_j,\phi_j)\cdot\mathbf{r_{ij}},\\
\label{eq:31_2}
    \mathbf{\alpha}_{ij} &=& R(\theta_j-\theta_i,\phi_j-\phi_i)\cdot\mathbf{r_{ij}}.
\end{eqnarray}
\end{subequations}
  
Therefore, the net field on macrospin $i$, expressed in the basis of $i$,  is
\begin{eqnarray}
\label{eq:32}
     \mathbf{H}_{\mathrm{d},ij}^{(i)}&=& \frac{1}{4\pi}\sum_{\tau_1,\tau_2}^\mathrm{ASI}\sum_j^\mathrm{U.C.} V_jM_{s,j} \Big[\frac{\mathbf{\alpha}_{ij}(\mathbf{\rho}_{ij}\cdot\mathbf{m_j})}{r_{ij}^5}\nonumber\\&&-\frac{R(\theta_j-\theta_i,\phi_j-\phi_i)\cdot\mathbf{m_{j}}}{r_{ij}^3}\Big] 
 \end{eqnarray}

 The last step to collapse the sums into a single Hamiltonian matrix is to incorporate a tight-binding approach. We invoke Bloch's theorem but we make the assumption that the phase between the complex amplitudes is solely given by the translation vectors between unit cells. This is the main approximation in our model and ensures that the resulting band structure is periodic within the FBZ. If macrospin-to-macrospin phases were to be included, then length scales smaller than the FBZ would be resolved, which is outside the model's scope. Therefore, we apply Bloch's theorem as
\begin{equation}
\label{eq:33}
    \mathbf{m}_j^{(\tau_1,\tau_2)} = \mathbf{m}_j e^{\Phi}=\mathbf{m}_j e^{-i\mathbf{a}_{ij}^{(\tau_1,\tau_2)}\cdot\mathbf{k}}\,
\end{equation}
and compute the 
dipole energy, 
\begin{equation}
    \mathbf{\mathcal{H}^{(i)}} =  \mu_0M_{s,i}\mathbf{m}_{i}^T\cdot\mathbf{H}_{\mathrm{d},ij}^{(i)}.
\end{equation}


Rewriting the energy as a function of the complex amplitudes $a$ and rescaling to units of frequency, we obtain the block Hamiltonian matrices
 \begin{subequations}
 \label{eq:34}
\begin{eqnarray}
\label{eq:34_1}
       \underline{\Omega}_{d}^{1,1} &=& \sum_{\tau_1,\tau_2}^\mathrm{ASI}e^\Phi\begin{bmatrix}
       0 & C_{12} & 0 \\
       C_{21} & 0 & C_{23} \\
       0 & C_{32} & 0 
       \end{bmatrix} \\
      \label{eq:34_2} 
      \underline{\Omega}_{d}^{1,2} & =& \sum_{\tau_1,\tau_2}^\mathrm{ASI}e^\Phi\begin{bmatrix}
       G_{11} & D_{12} & 0 \\
       D_{21} & G_{22} & C_{23} \\
       0 & D_{32} & G_{33}
       \end{bmatrix},
 \end{eqnarray} 
\end{subequations}
where, 
 \begin{subequations}
\begin{eqnarray}
\label{eq:35_1}
     C_{ij} &=& \frac{3}{|\mathbf{r}_{ij}|^5}[\alpha_1\rho_1+i(\alpha_1\rho_2+\alpha_2\rho_1)- \alpha_2\rho_2]\nonumber\\&&- \frac{1}{|\mathbf{r}_{ij}|^3}[R^{(1,1)}-R^{(2,2)}+ i(R^{(1,2)}+R^{(2,1)})],\\
     \label{eq:35_2}
   D_{ij} &=& \frac{3}{|\mathbf{r}_{ij}|^5}[\alpha_1\rho_1+i(\alpha_1\rho_2-\alpha_2\rho_1) + \alpha_2\rho_2]\nonumber\\&&- \frac{1}{|\mathbf{r}_{ij}|^3}[R^{(1,1)}+R^{(2,2)}-i(R^{(1,2)}-R^{(2,1)})],\\
   \label{eq:35_3}
    G_{ij} &=& -\frac{3}{|\mathbf{r}_{ij}|^5}[\alpha_3\rho_3] -\frac{2}{|\mathbf{r}_{ij}|^5}R^{(3,3)},
     \end{eqnarray}
     \end{subequations}
and we have used a slightly shorthand notation where $\rho_{i,j}=(\rho_1,\rho_2,\rho_3)$, $\alpha_{i,j}=(\alpha_1,\alpha_2,\alpha_3)$, $R$ represent the components of the $3\times3$ matrix $R(\theta_j-\theta_i,\phi_j-\phi_i)$, and the sums over the ASI modify the phase $\Phi$.    

\section{Validation}

G\ae{}nice is implemented in MATLAB and it can be obtained from http://doi.org/10.17605/OSF.IO/YUNHD as well as a script that reproduces the results presented below.
\begin{figure*}[t]
\centering \includegraphics[width=7in]{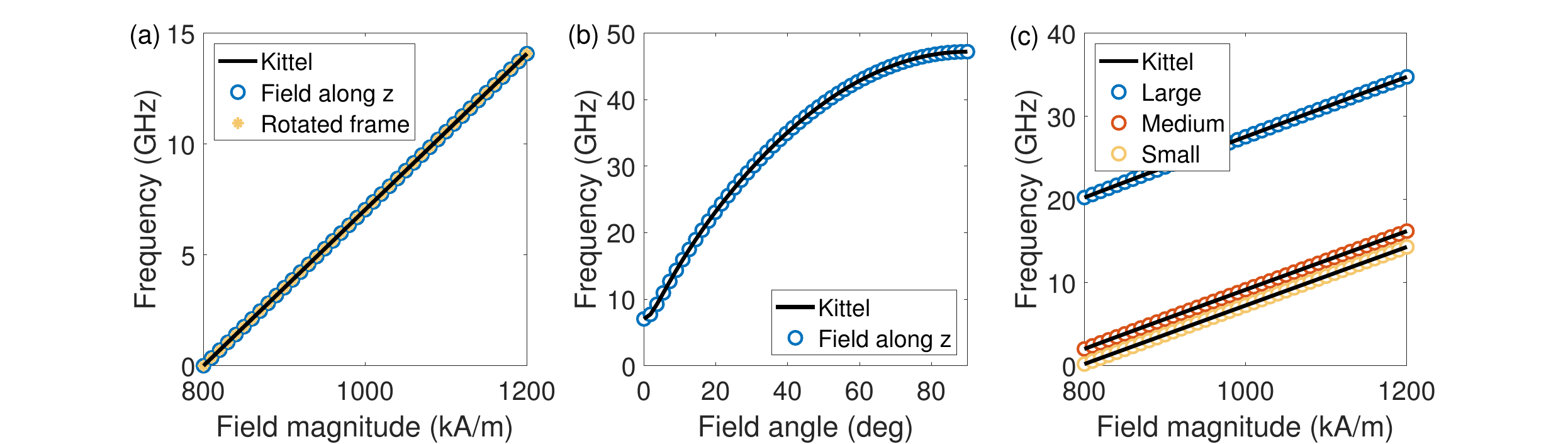}
\caption{ \label{figKittel} Comparison of numerical computation of ferromagnetic resonance (FMR) and Kittel's equation to validate the geometry and implementation of the rotation matrix. (a) Field-dependent FMR of a perpendicularly magnetized easy-plane ferromagnet. The blue circles represent calculations for a magnet in the $x-y$ plane and magnetized along the $z$ direction. The calculations agree with Kittel's equation \eqref{eq:kittel}. The {gold} asterisks are obtained when the magnet and the magnetization are oriented along $\theta_d=\theta=33$~deg and $\varphi_d=\varphi=233$~deg. The field is perpendicular to this orientation, and we recover the same field-dependent frequency. (b) Angle dependence FMR of an in-plane magnetic saturated at $H=1,000$~kA/m. Both numerical computations (blue circles) and Kittel's equation~\eqref{eq:kittel_angle} agree. (c) Validation of the anisotropy field implementation. Field-dependent FMR for magnets of different sizes (colored symbols) and the corresponding Kittel's equation \eqref{eq:kittel_size}. }
\end{figure*}

\subsection{Local fields}

The validity of the implementation of local fields can be verified my means of the Kittel equation. For the purposes of the model presented here, it is imperative to verify the field magnitude and angle dependent ferromagnetic resonance as well as its independence from the coordinate system.

We first model a circular thin-film which can be considered as a single macrospin due to the fact that only the hard axis contributes to the demag tensor, i.e., $D_3=1$ and $D_1=D_2=0$. We use a saturation magnetization of  $M_s=800$~kA/m. 
Kittel's equation as a function of the magnetic field amplitude $H$ is thus:
\begin{equation}
\label{eq:kittel}
    \omega = \gamma\mu_0M_s\left(\frac{H}{M_s}-1\right),
\end{equation}
valid for $H>M_s$. To model this scenario, we use an applied field oriented along the $z$-axis with magnitude in the range $800$~kA/m~$<H<1,200$~kA/m. The magnetization is parallel to the applied field so that $\theta_m=0$ and $\varphi_m=0$. The magnetic film must be rotated so that the hard axis is also oriented along the $z$-axis. In other words, $\theta_d=\pi/2$ and $\varphi_d=0$. The numerical results are shown in Fig.~\ref{figKittel}(a) by blue circles. The solution of the Kittel equation is shown at the top by a solid black line. The agreement is within numerical error ($<4\times10^{-15}$).

Validation of the rotation matrix is achieved by computing the same field dependence when the magnetization, external field, and nanomagnet are rotated by arbitrary polar and azimuth angles. For example, selecting a rotation $\theta=33$~deg and $\varphi=233$~deg, we recover the correct solution within numerical error ($<9\times10^{-15}$), shown in Fig.~\ref{figKittel}(a) by gold asterisks.

We now set the external field magnitude to $H=1,000$~kA/m and vary its angle, $\theta_0$. The frequency as a function of angle is obtained from the Kittel equation expressed as
\begin{equation}
\label{eq:kittel_angle}
    \omega = \gamma\mu_0\sqrt{H_i\left(H_i+M_s\cos^2{(\theta_0)}\right)},
\end{equation}
where $H_i$ is the internal magnetic field magnitude obtained by solving the magnetostatic equations
\begin{subequations}
\begin{eqnarray}
    (H_i + M_s)\cos{\theta_i} &=& H\cos{(\theta_0)},\\
    H_i\sin{(\theta_i}) &=& H\sin{(\theta_0)}.
\end{eqnarray}
\end{subequations}

The magnetization vector is oriented along the internal magnetic field angle for a saturating field, $\theta_m=\theta_i$ and $\varphi_m=0$. We define $\theta_0=0$ as the out-of-plane component so that $\theta_d=\pi/2$ and $\varphi_d=0$ for all cases. The results shown in Fig.~\ref{figKittel}(b) further validate the implementation of the external field and demag fields.

Finally, we vary the size of the magnetic element so that all three demagnetizing factors are computed. We consider three different oblate spheres: ``Large'' ($10,000~\mathrm{nm}\times 1,000~\mathrm{nm}\times5$~nm), ``Medium'' ($1,000~\mathrm{nm}\times 100~\mathrm{nm}\times5$~nm), and ``Small'' ($100~\mathrm{nm}\times 10~\mathrm{nm}\times5$~nm). The field is once again considered to be oriented along the $z$-axis and its magnitude is varied between $800$~kA/m~$<H<1,200$~kA/m. The frequency dependence as a function of field is given by Kittel's equation
\begin{equation}
\label{eq:kittel_size}
    \omega = \gamma\mu_0M_s\sqrt{\left(\frac{H}{M_s}+D_1-D_3\right)\left(\frac{H}{M_s}+D_2-D_3\right)}.
\end{equation}

The results shown in Fig.~\ref{figKittel}(c) validate the demagnetization field and its nanomagnet-dependent implementation because all three nanomagnets are concurrently simulated. This also shows that G\ae{}nice can be used as a tool to quickly compute FMR for an ensemble of uncoupled nanomagnets.

\subsection{Nonlocal field: exchange}

The dynamic contribution of the exchange energy introduces the splitting of the resonant frequencies within a single nanomagnet. As a test case, we set a stadium-shaped nanomagnet with dimensions $l=280$~nm, $w=100$~nm, and $t=10$~nm. 
The nanomagnet is oriented along the $x$ axis, $\theta_d=\pi/2$ and $\varphi_d=0$.

\begin{figure}[t]
\centering \includegraphics[width=2.5in]{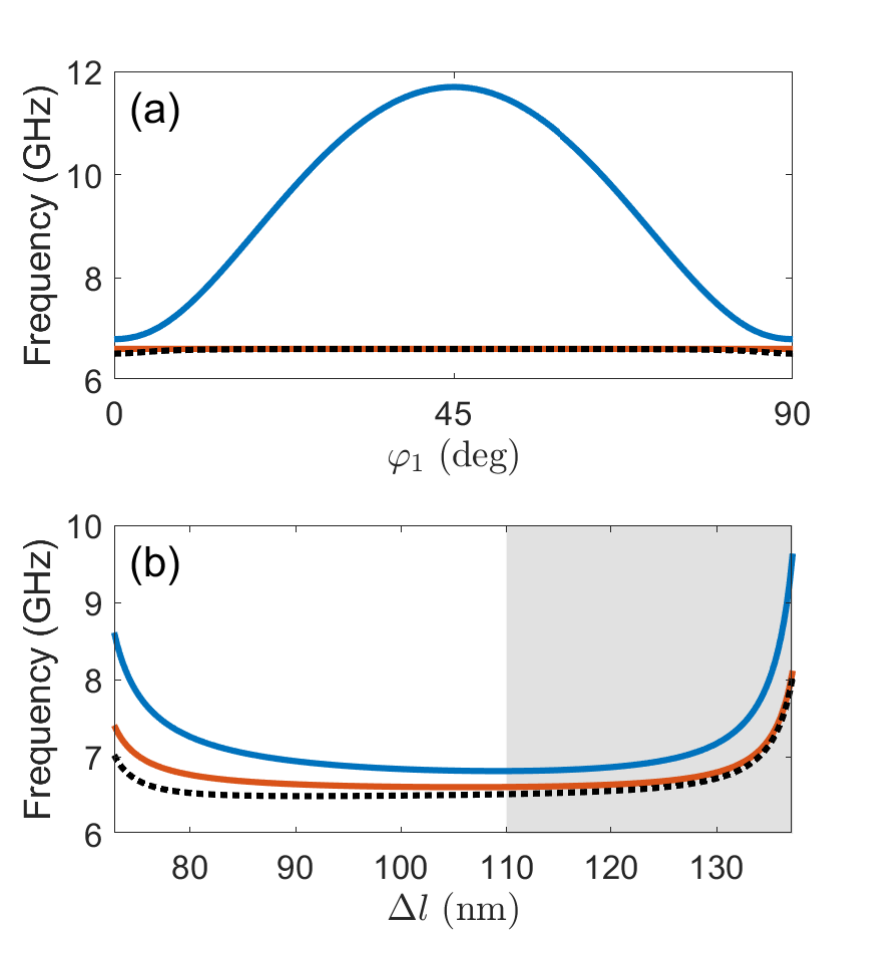}
\caption{ \label{fig:exchange} Frequencies computed by adding the exchange Hamiltonian. (a) A single macrospin is azimuthally rotated, resulting in a sizeable variation of the frequency in one band. The minima occur at $\varphi_1=0$~deg and $\varphi_1=90$~deg, consistent with a dynamic coupling mediated only by the magnetization's $z$ component. (b) Frequency variation as a function of $\Delta l$, showing divergence as either the bulk or edge volumes tend to zero. The frequencies in the vicinity of the default value $\Delta l=(2l-w)/4$ (the transition between the white and gray areas) {are} approximately constant.  Each band is displayed in different colors for clarity.}
\end{figure}
\begin{figure}[t]
\centering \includegraphics[width=2.5in]{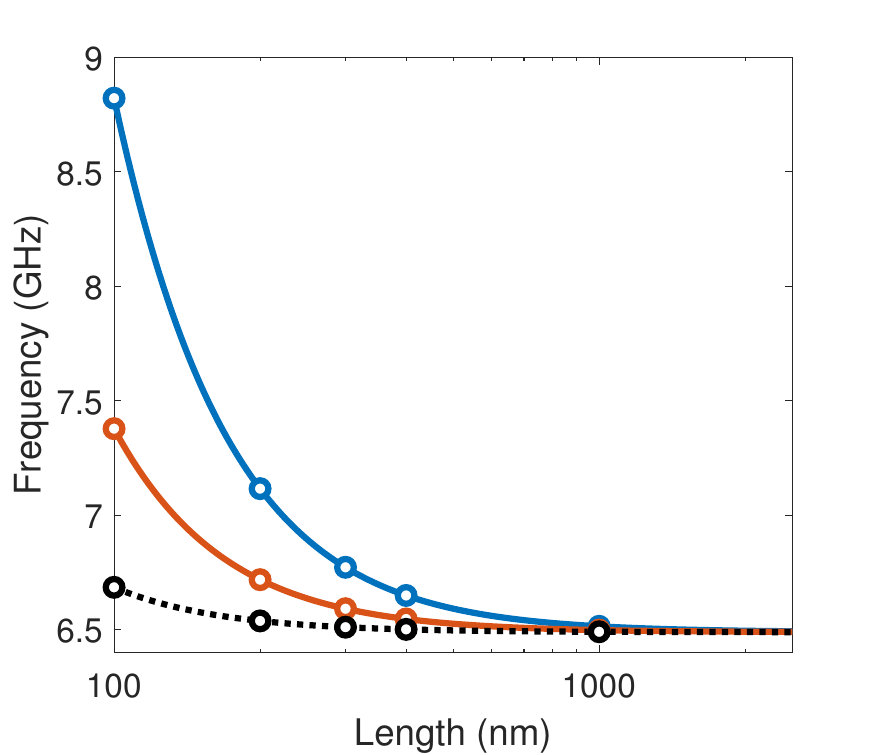}
\caption{ \label{fig:exchange_length} Frequency dependence on the nanomagnet's size. The aspect ratio is maintained in the calculations so that the length is a representative metric. The solid lines are frequencies obtained from computations on a single nanomagnet. The circles are obtained from a single computation including five non-interacting nanomagnets. Each band is displayed in different colors for clarity. }
\end{figure}
We first explore the effect of the magnetization's relative angles. For this, we set the magnetization parallel to the nanomagnet orientation, and we vary the azimuth angle of the magnetization at one extremum, $\varphi_1$. The computed frequencies are shown in Fig.~\ref{fig:exchange}, where different colors and dashed curves were used for each branch for clarity. One frequency branch exhibits a sinusoidal variation, consistent with one magnetization being rotated and modifying the exchange energy. The maximum occurs at $45$~deg, implying that the maximum exchange contribution occurs when the adjacent magnetization vectors dynamically couple in both $x$ and $y$. Indeed, when $\varphi_1=0$~deg or $90$~deg, the $z$ components of the magnetization vectors are coupled, leading to identical energy contributions to the eigenmodes. Note that this is different than the static exchange energy computed in Eq.~\eqref{eq:18}.

We now explore the influence of the bulk and edge volume ratios. From Eq.~\eqref{eq:26}, the exchange energy diverges as either the bulk or edge volume tends to zero. This is expected because of the underlying assumption that the nanomagnet is separated in three macrospins. In other words, such a divergence has no physical origin. The frequencies computed as a function of $\Delta l$ when all the magnetization vectors are aligned with the nanomagnet are shown in Fig.~\ref{fig:exchange}(b). Clearly, the frequencies diverge when the bulk and edge volumes tend to zero towards the left and right extrema of the figure, respectively. The frequencies are relatively constant close to the default distance $\Delta l=(2l-w)/4=110$~nm.

As the size of the nanomagnet increases, the effect of the exchange interaction in the frequencies must necessarily decrease insofar as the nanomagnet is split {into} three macrospins. We compute this test scenario by locking $\Delta l$ to its default value and varying the nanomagnet's size. We maintain the aspect ratio of the nanomagnet so that the length is representative of the nanomagnet's volume. The results are shown by solid and dashed curves in Fig.~\ref{fig:exchange_length}, where the x axis is shown in natural logarithmic scale and the colors represent different branches for clarity. As expected, the frequency splitting diminishes as the nanomagnet's size increases.

A final test for the exchange interaction, is to verify that its implementation is independent of the number of nanomagnets. For this, we specify five nanomagnets with lengths $100$, $200$, $300$, $400$, and $1000$~nm and dimensions consistent with the aspect ratio of the test case considered in this section. The resulting eigenvalue problem requires solving for a matrix of dimension $30\times30$. The frequencies are shown by circles in Fig.~\ref{fig:exchange_length}, color-coded according to the branches of the single-nanomagnet calculations. We note that the frequencies are not automatically sorted for each nanomagnet: only the computation of the eigenvectors can return such type of sorting which is not currently computed in our implementation, as discussed in the conclusions. In this case, the frequencies were manually sorted. The results are in agreement with the calculations done for each nanomagnets, validating that the exchange interaction is nonlocal but intrinsic to each nanomagnet, i.e, there is no coupling between nanomagnets.

\begin{figure}[t]
\centering \includegraphics[width=2.5in]{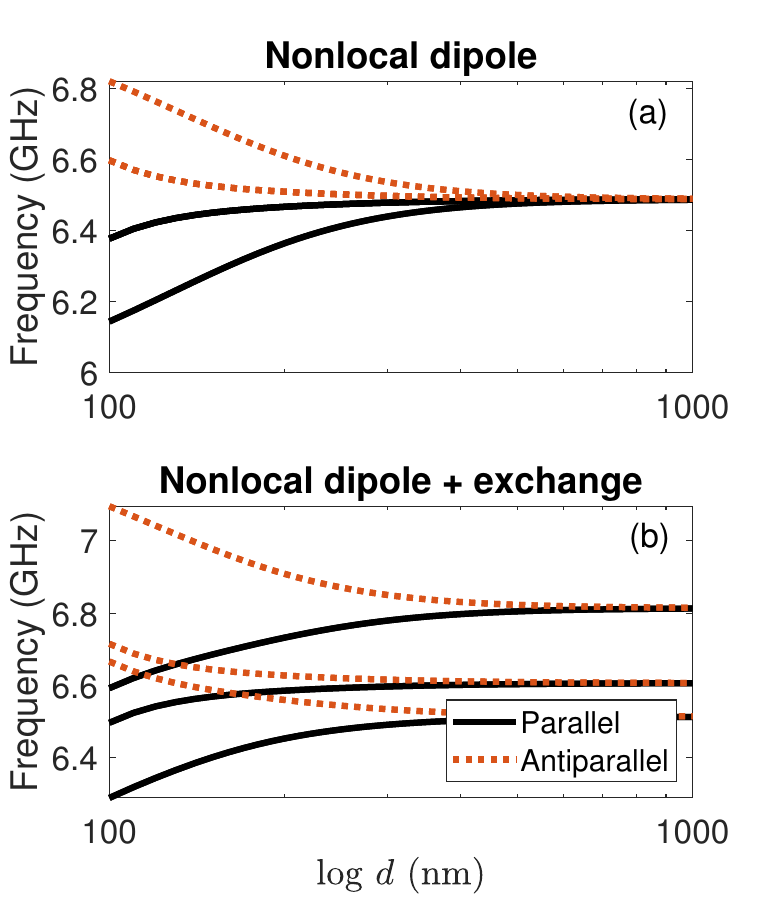}
\caption{ \label{fig:dipole_dist} Frequencies as a function of distance between two identical nanomagnets interacting via (a) nonlocal dipole field and (b) both nonlocal dipole field and exchange interaction within each nanomagnet. The magnets are parallel to one another, and we distinguish the relative magnetization being parallel (solid black curves) and anti-parallel (dashed red curves). In all cases, the nonlocal dipole field becomes negligible at large distances and the bands become degenerate, as expected for non-interacting identical nanomagnets. }
\end{figure}
\begin{figure}[t]
\centering \includegraphics[width=3in]{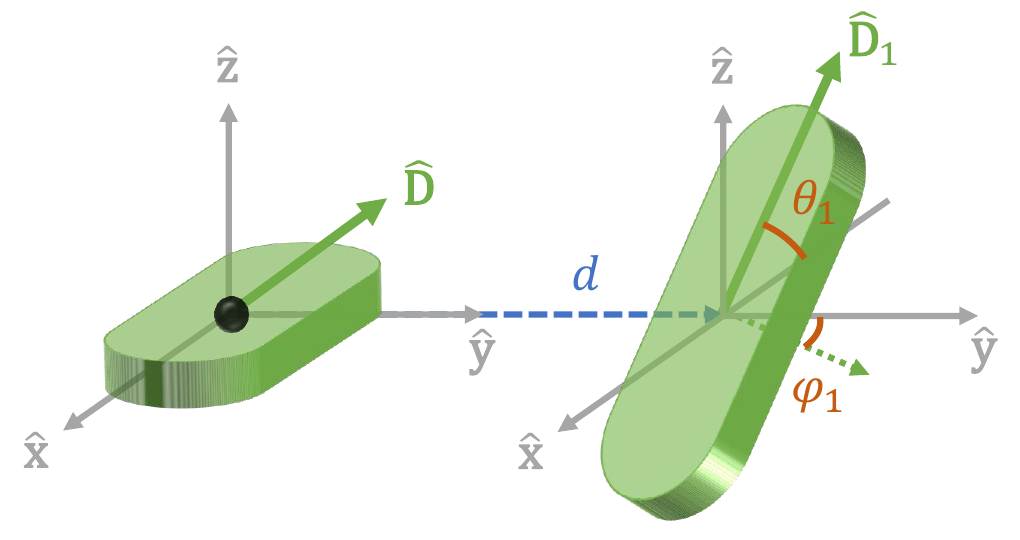}
\caption{ \label{fig:stadium_rotation_schem} Geometrical variations between two interacting nanomagnets. The nanomagnet with director $\hat{D}_1$ is located at a distance $d=300$~nm along the $\hat{y}$ direction. The director's orientation is varied by the polar and azimuth angles $\theta_1$ and $\varphi_1$, respectively. }
\end{figure}
\begin{figure}[t]
\centering \includegraphics[trim={.3in 0in 0.6in 0in}, clip, width=3.3in]{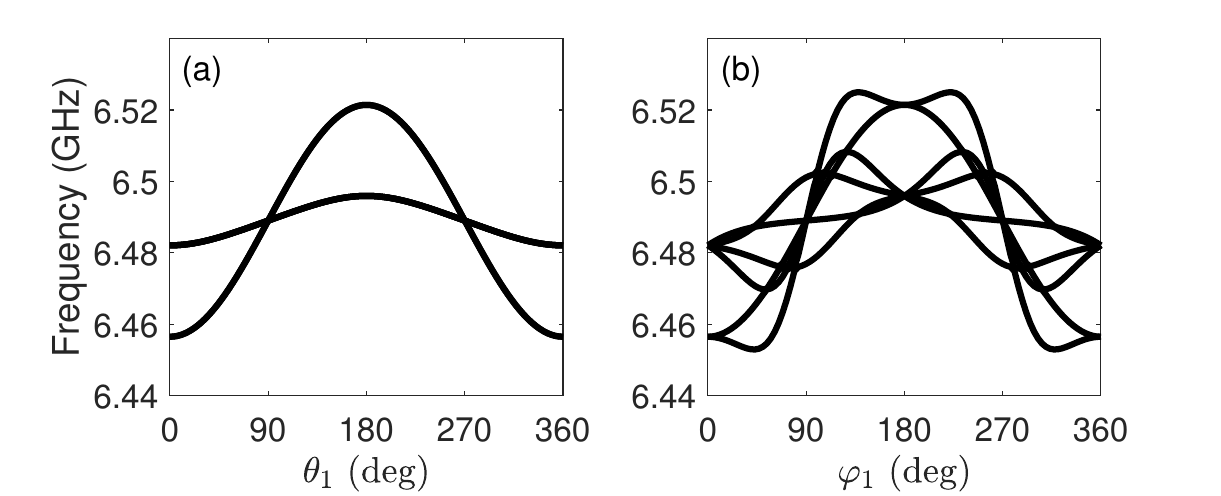}
\caption{ \label{fig:stadium_rotation} Computed frequencies for a pair of interacting nanomagnets when one of the nanomagnets is rotated about (a) the polar angle $\theta_1$ and (b) the azimuth $\varphi_1$. The angles are shown in the schematic Fig.~\ref{fig:stadium_rotation_schem}.}  
\end{figure}

\subsection{Nonlocal field: dipole}

{We now investigate 
the interaction between two identical nanomagnets of dimensions $l=280$~nm, $w=100$~nm, and $t=10$~nm, as used in the previous section. 
We focus here on collective excitation, so that $|\mathbf{k}|=0$ and the phase contributions in Eqs.~\eqref{eq:34} simplify to 1.

The first test ensures that the nonlocal dipole field's strength depends on the distance between the nanomagnets. For this, we consider a varying distance $d$ along the $y$ axis ranging from $100$~nm to $1,000$~nm. The computed frequencies considering only nonlocal dipole fields are shown in Fig.~\ref{fig:dipole_dist}(a). The relative magnetization orientation is parallel for the solid black curves and antiparallel for the dashed red curves. In both cases, the modes are degenerate at long distances. This is a clear indication that nonlocal dipole field does not affect the internal modes of non-interacting (or weakly interacting) nanomagnets. Modes are visibly split under a distance of $\approx400$~nm. Red and blue-shifts are observed for the parallel and antiparallel cases, respectively, in agreement with the static dipole energy for each. Including exchange energy,shown in Fig.~\ref{fig:dipole_dist}(b), 
naturally leads to larger split bands because of the additional energy. 
As expected in all cases, the modes converge towards degenerate values at large distances indicating a negligible interaction mediated by the nonlocal dipole field.

We next explore the frequency dependence on the relative orientation between the two nanomagnets. For this, we consider a nanomagnet located at the global origin of the Cartesian coordinate and  {a} second nanomagnet located at a distance of $d=300$~nm in the $y$ direction with a varying unit vector $\hat{\mathbf{D}_1}$. We consider both polar and azimuth rotations parametrized by the angles $\theta_1$ and $\varphi_1$, respectively, as shown Fig.~\ref{fig:stadium_rotation_schem}. Note that these angles are measured relative to the orientation of the fixed nanomagnet.

The computed frequencies are shown in Fig.~\ref{fig:stadium_rotation}. The frequency variation as a function of the polar angle $\theta_1$ is shown in (a). In these computations, we disabled the exchange interaction to focus on the symmetry of the static dipole field. 
There is a modest change in the frequency that is maximal at $\theta_1=180$~deg. The symmetry is also consistent with the fact that $90$~deg and $270$~deg are degenerate. The frequency variation as a function of the azimuth angle $\varphi_1$ is shown in (b). There are again clear symmetries consistent with the rotation of the nanomagnet despite the increased number of modes originating from the non-collinear magnetization orientations. Notably, at $90$~deg and $270$~deg, the rotated nanomagnet is perpendicular to the fixed nanomagnet and the spacing is just $110$~nm. Strong variations are observed close to these conditions. As expected, the computed frequencies are periodic for both $\theta_1$ and $\varphi_1$.

This concludes the verification of the static dipole field, which follows the qualitative expectations of decay with distance and symmetries due to different types of rotations.

\subsection{Band structure}


\subsubsection{Nanomagnet chain}

A one-dimensional chain of nanomagnets is modeled by a single nanomagnet with dimensions $l=280$~nm, $w=100$~nm, and $t=10$~nm subject to a translation vector $\mathbf{a}_1$ oriented at an azimuth $\varphi_a$ and lattice constant $|\mathbf{d}|=300$~nm. Magnons with wave vectors $\mathbf{k}$ oriented at an azimuth $\varphi_k$ are computed, as shown schematically in Fig.~\ref{fig:stadium_chain_schem}. 


The magnon dispersion is computed for cases where we set the wavevector parallel to the $x$-axis and rotating the translation vector. 
In other words, $\mathbf{a}_1=d\hat{\mathbf{a}}_1=d[\cos{(\varphi_a)}\hat{x}+\sin{(\varphi_a)}\hat{y}]$. 
The resulting dispersion relations as a function of $k=2\pi/d(\mathbf{k}\cdot\hat{\mathbf{a}}_1)$ within the first Brillouin zone for selected azimuths are shown in Fig.~\ref{fig:stadium_rotation_k}(a). Three bands are observed because a single nanomagnet composes the unit cell of the chain. The band structures show a pronounced periodic behavior in the FBZ that is symmetric relative to $\varphi_a=90$~deg, stemming from the product $\mathbf{k}\cdot\mathbf{a}_1$. This symmetry validates the implementation of the sums performed in Eqs.~\eqref{eq:34}. 
It is also shown in the mid panel of Fig.~\ref{fig:stadium_rotation_k}(a) that the band structure perpendicular to the chain orientation at $\varphi_a=90$~deg, is flat. This is because the phase in Eqs.~\eqref{eq:34} is exactly zero when the translation vector and wavevector are perpendicular. 
\begin{figure}[b]
\centering \includegraphics[width=3in]{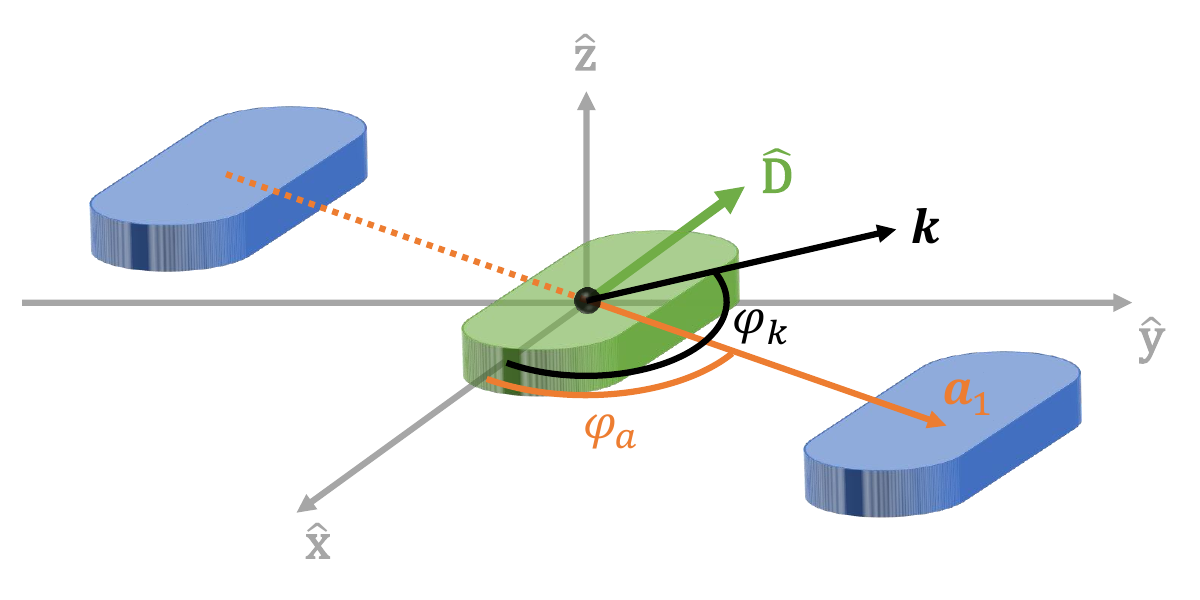}
\caption{ \label{fig:stadium_chain_schem} A chain of nanomagnets is modeled by a single nanomagnet (green) upon which the nonlocal dipole field from an infinite chain of nanomagnets (blue) acts. The chain can be defined along an arbitrary in-plane direction by setting the translation vector $\mathbf{a}_1$ oriented at an azimuth $\varphi_1$. The magnon dispersion can be computed for arbitrary in-plane wavevectors $\mathbf{k}$ given the azimuth $\varphi_k$. }
\end{figure}
\begin{figure}[t]
 \centering \includegraphics[width=3in]{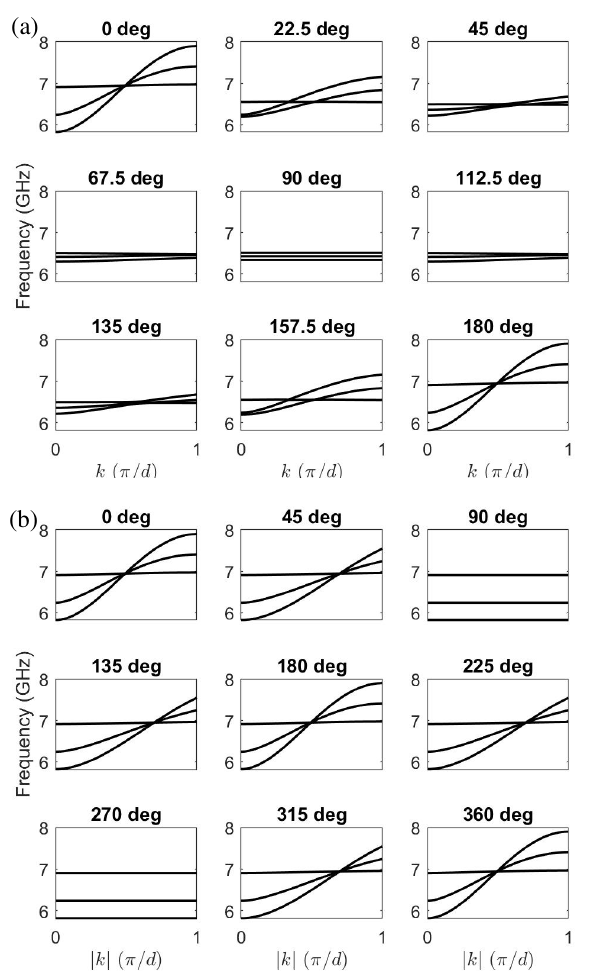}
 \caption{ \label{fig:stadium_rotation_k} Dispersion relation for a 1D nanomagnet chain upon (a) setting the wavevector along $\hat{x}$ and rotating $\mathbf{a}_1$ and (b) setting  $\mathbf{a}_1$ along the $\hat{x}$ direction and rotating the wavevector. The bands exhibit the most changes when the wavevector and translation vector are parallel and are flat when these are orthogonal. This is in agreement with our tight-binding definition of the phase in Eqs.~\eqref{eq:34}.}  
 \end{figure}

 \begin{figure}[t]
 \centering \includegraphics[trim={0in 0.2in 0in 0in}, clip, width=2.5in]{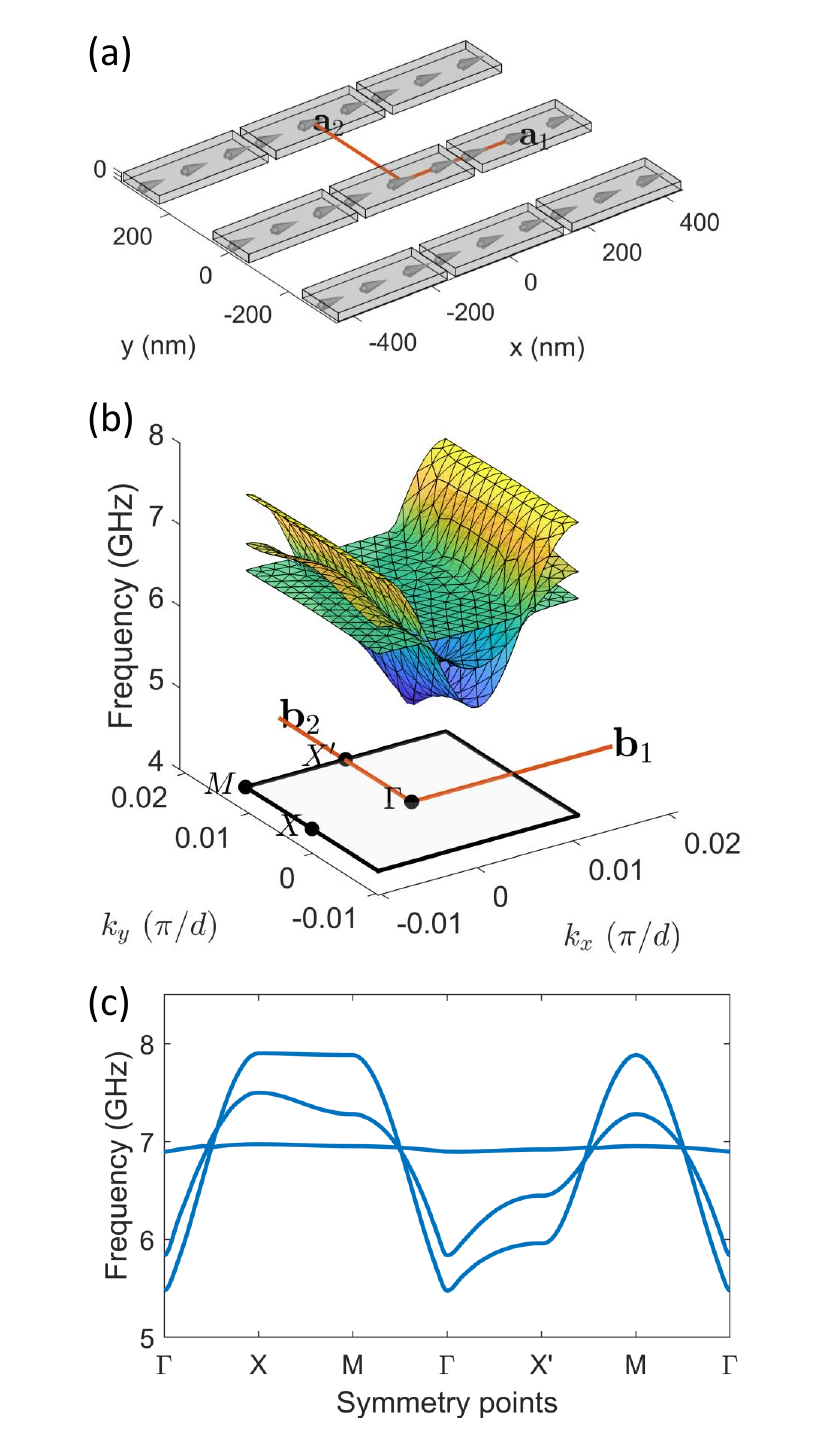}
 \caption{ \label{fig:stadium_2d} (a) G\ae{}nice representation of the {2D array of nanomagnet chains}, depicting the translation vectors. (b) Resulting band structure where the FBZ is shown under the band structure. The FBZ is directly computed by G\ae{}nice from the translation vectors and the high-symmetry points are also identified. The band structure is obtained by performing a Delaunay triangulation over the FBZ and evaluating the resulting wavevectors. The color scale represents the frequency and it is also shown in the vertical axis. (b) Irreducible path in the FBZ exhibiting the asymmetry of this geometry as well as the periodicity achieved by the tight-binding method.}  
 \end{figure}
 \begin{figure*}[t]
 \centering \includegraphics[width=7in]{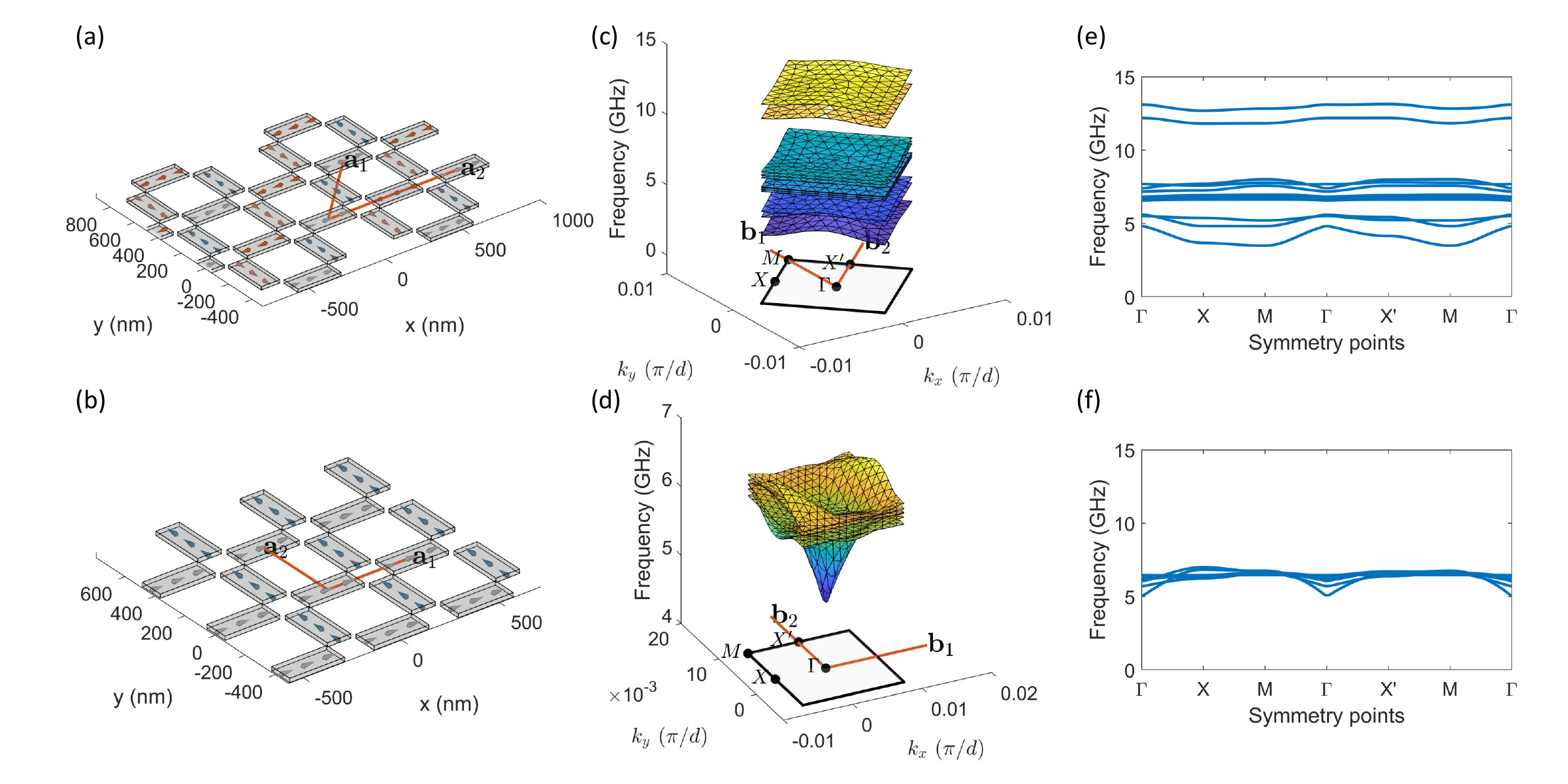}
 \caption{ \label{fig:stadium_square} G\ae{}nice representation of the square ice geometry for the (a) vortex and (b) remanent states. The respective band structure for each case in shown in (c) and (d) while the irreducible path in the FBZ are shown in (e) and (f).}  
 \end{figure*}

It is also possible to consider another case of $\varphi_a\neq\varphi_k$. We set the translation vector along the $x$ axis, i.e., $\mathbf{a}_1=d\hat{x}$ and we vary $\varphi_k$, such that $\mathbf{k}$= $|\mathbf{k}| [\cos{(\varphi_k)}\hat{x}+\sin{(\varphi_k)}\hat{y}]$. 
The dispersion relations for $k=|\mathbf{k}|$ up to the FBZ are shown in Fig.~\ref{fig:stadium_rotation_k}(b). As for Fig.~\ref{fig:stadium_rotation_k}(a), the expected symmetries are respected, e.g. the band structure perpendicular to the chain orientation is flat; see $\varphi_k=90$ and $270$~deg, for the same reasons outlined above.. Note that here we extend the rotation of $\varphi_k$ to a full cycle.

\subsubsection{Multiple interacting chains}

We now calculate the band structure of interacting nanomagnet chains. Each nanomagnet is oriented at $\phi_d=0$ with respect to the x-axis and the 
array is generated from the single nanomagnet unit cell due to translation vectors $\mathbf{a}_1$ and $\mathbf{a}_2$ with a lattice constant of $d=300$~nm. A visualization of this configuration is produced by G\ae{}nice to ensure the correct geometry definition, shown in Fig.~\ref{fig:stadium_2d}(a). The magnon band structure is computed by an automatic determination of the FBZ and its subsequent Delaunay triangulation to produce an array of wavevectors $\mathbf{k} = k_x\hat{x}+k_y\hat{y}$. This feature allows to optimally map the FBZ and produce band surfaces
, as shown in Fig.~\ref{fig:stadium_2d}(b). 

As for the 1D nanomagnet chain, the FBZ also shows three bands because the unit cell consists of a single nanomagnet. 
By examining the band structure depicted in Fig.~\ref{fig:stadium_2d}(b), the frequencies calculated along $k_x\hat{x}$ exhibit pronounced variations while it is predominantly flat along $k_y\hat{y}$.  This is consistent with our tight-binding approach whereby the dynamic 
dipole coupling depends on the gap distance between nanomagnets. 
 \begin{figure*}[t]
 \centering \includegraphics[width=7in]{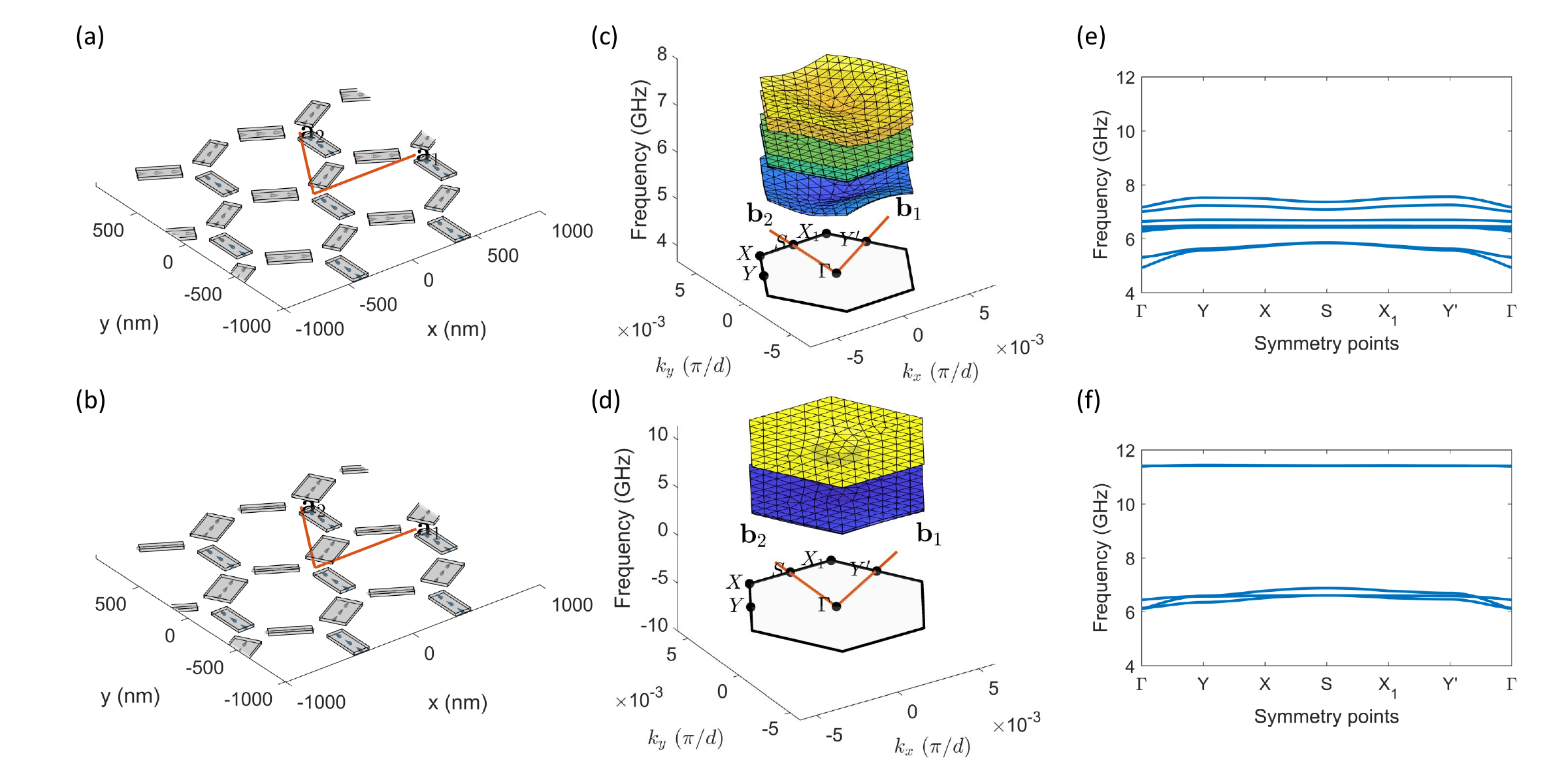}
 \caption{ \label{fig:stadium_kagome} { G\ae{}nice representation of the Kagome ice geometry for (a) identical nanomagnets and (b) anisotropy modified nanomagnets. The respective band structure for each case in shown in (c) and (d) while {the} irreducible path in the FBZ exhibiting are shown in (e) and (f).}  }
 \end{figure*}

The irreducible path in the FBZ can be also directly computed in G\ae{}nice. We observe that the band structure is different when the path is taken towards the $X$ and $X'$ points. This is because the array is asymmetric, {such} that the dipole field is different along $\hat{x}$ and $\hat{y}$ directions.

\subsubsection{Square ice}

We now use G\ae{}nice to compute the magnon band structure for square ASI, where four nanomagnets are placed around a vertex, each at an angle of 90 degrees to one another and equidistant from the vertex. We maintain the previously used nanomagnet dimensions $l=280$~nm, $w=100$~nm, and $t=10$~nm but now set a center-to-center distance of $d=430$~nm. For simplicity, we investigate the band structure for states where the magnetization is in a homogeneous (onion) state. Edge bending in the magnetization state leads to $S$ and $C$ states that are known to modify the band structure~\cite{Iacocca2016}.

We investigate both the vortex (type-I) and remanent (type-II) configuration. The vortex state has four nanomagnets in the unit cell and is defined by the translation vectors  $\mathbf{a}_1 = 2\hat{x}$ and $\mathbf{a}_2 = \hat{x}+\hat{y}$. The remanent state has two nanomagnets in the unit cell and is defined by $\mathbf{a}_1 = \hat{x}$ and $\mathbf{a}_2 = \hat{y}$. These configurations are shown in Fig ~\ref{fig:stadium_square}(a) and (b), respectively. 

The band structures in the FBZ are shown in Fig.~\ref{fig:stadium_square}(c) and ~\ref{fig:stadium_square}(d), for the vortex and remanent states, respectively. 
By dividing the nanomagnet into three macrospins, the vortex state has twelve bands. {The band structure exhibits little dispersion, which may be expected by the fact that the stray fields are largely compensated in a type-I configuration. The two high-frequency bands are bulk modes while the low-frequency bands are edge modes, {as shown previously~\cite{Iacocca2016}}. This band separation is clearly seen in {the irreducible path shown in } Fig.~\ref{fig:stadium_square}(e), exhibiting a band-gap of about $\approx5$~GHz.

In the remanent state, there are six bands. In this case, the bands are very close together, with a visible dip at the $\Gamma$ point. It is also evident that the band structure is skewed, which is a consequence of the likewise skewed static dipole field in this configuration.

The results in this section {are in agreement with previous calculations~\cite{Iacocca2016} demonstrating the reconfigurability of the magnon band structure for square ices.} However, the improved dipole field implementation in G\ae{}nice showcases more subtleties in the band structure as well as asymmetries that could {in principle indicate directional magnon propagation, as recently surmised in a combined experimental and micromagnetic study~\cite{Lendinez2023}}. We also emphasize that we have only explored here the onion state, but it is well-known that the magnetization tilts at the edges of the nanomagnets due to stray fields. This adds an additional degree of freedom for tuning the band structure. 

\subsubsection{Kagome ice}

We now explore the band structure for Kagome ASI. The Kagome unit cell comprises three nanomagnets with lattice constant $d = 800$~nm which we define as twice the radius of the circle in which the hexagonal structure is embedded. Considering the center of the triad of nanomagnets as the origin, we define the translation vectors $\mathbf{a}_1=\hat{x}$ and $\mathbf{a}_2=(1/2)\hat{x}+(1/\sqrt{3}+1/4)\hat{y}$.

We consider two cases: {a ``regular'' Kagome ice where the nanomagnets have identical dimensions $l=280$~nm, $w=100$~nm, and $t=10$~nm; and an anisotropy modified Kagome ice inspired by the work by T. Dion et al.~\cite{Dion2019}, where we use three different widths $w=100$~nm, $w=180$~nm, and $w=60$~nm for the nanomagnets in the unit cell. The geometries are} 
shown in Fig.~\ref{fig:stadium_kagome}(a) and  Fig.~\ref{fig:stadium_kagome}(b). {In both cases, the array is in a degenerate ground state where the unit cell triad has a 2-in/1-out vertex.}

In the ``regular'' Kagome ice we find a modest band structure with all nine bands contributing to the band structure, shown in Fig.~\ref{fig:stadium_kagome}(c). However, the anisotropy modified Kagome ice exhibits only four bands, as shown in Fig.~\ref{fig:stadium_kagome}(d) with other five softened to exactly zero. In the context of our framework, a zero-frequency band entails a real, evanescent solution. The bands in the irreducible paths in Figs.~\ref{fig:stadium_kagome}(e) and (f) further confirm that the bands are relatively flat in all cases. An important distinction is the anisotropy modified Kagome ice exhibits band-gaps which is consistent with the different FMR for each nanomagnet, i.e., different demagnetization factors. {While this is certainly not an in-depth investigation of the frequency response of anisotropy modified Kagome ices, it showcases the functionality of G\ae{}nice to compute the band structure of relatively complex geometries with ease.}

\section{Conclusions}

We have presented G\ae{}nice, a computational tool to compute the dispersion relation of arbitrary artificial spin ice geometries. The theoretical framework of G\ae{}nice relies on the excitation of small-amplitude perturbations and produces the dispersion relation by computing both static and dynamic dipole contributions to the Hamiltonian matrices. Our framework also relies on a tight-binding approach to ensure {the} periodicity of solutions within the FBZ, which composes the main simplification of the model.

G\ae{}nice can be also used for FMR computations of relatively complicated geometries. For example, G\ae{}nice has been recently applied for square ices based on trilayers and exhibited remarkable agreement with experiments and micromagnetic simulations of field-dependent FMR~\cite{Dion2023}. 
Because both the exchange and dipole interactions can be toggled, G\ae{}nice can be easily be used to study the FMR of ensembles of interacting or non-interacting nanoparticles and extended to 3D structures.

There are three main limitations to G\ae{}nice in its current form. First, the computations are accurate for nanoparticles and nanomagnets because of the assumption of three macrospins. Larger nanomagnets possess higher degrees of freedom that will reduce the relative energy contributions. Therefore, G\ae{}nice is likely to overestimate the frequency split when nanomagnets are brought {very} close together. A possible solution to this issue is to further split the nanomagnets into more macrospins, with the caveat that the number of macrospins should be kept to a minimum to maintain a computational advantage over micromagnetic simulations. Another way to solve this issue is to compute the energy of the system to actively modify the magnetization's edge bending due to stray fields, as recently shown in Ref.~\cite{Saccone2023}. Second, the wavefunctions are not currently computed. As is well known, the linearization of the Hamiltonian leads to wavefunctions with large errors. This limitation will be resolved in future work. Third, a 3D band structure is not currently supported. However, the basic framework is written and a generalization in 3D will compose a simple expansion of the dipole phases in the tight-binding approximation, a method that is well-known in solid-state physics.

\begin{acknowledgments}
This material is based upon work supported by the National Science Foundation under Grant No. 2205796. AR and EI acknowledge support from the UCCS Committee on Research and Creative Works (CRCW). VM and EI acknowledge the Department of Physics and Energy Science at UCCS for the use of their facilities and equipment. This work was supported by the Royal Academy of Engineering Research Fellowships, awarded to JCG. JCG was supported by EPSRC grant EP/X015661/1. Work by OGH at Argonne was supported by the US Department of Energy, Basic Energy Sciences Division of Materials Sciences and Engineering.
\end{acknowledgments}

\section*{Data Availability Statement}

The data that support the findings of this study are openly available in Open Science Framework (OSF) at http://doi.org/10.17605/OSF.IO/YUNHD.

\appendix
\section{Derivation of exchange energy}
\label{app:exchange}

To compute the exchange energy, we use a simple quasi-one-dimensional spin chain model to estimate the energy along the chain and relate it to the nanomagnet's regions and their volume. Consider a chain of length $l$ where the magnetization vector is linearly rotated, so that
\begin{equation}
\label{eq:17}
    \mathbf{m} = \cos{(k_0x)}\hat{\mathbf{x}}+\sin{(k_0x)}\hat{\mathbf{y}}.
\end{equation}

It can be shown that the exchange energy is given by $E_\mathrm{ex}=AVk_0^2+E_0$, where $A$ is the exchange constant in units of pJ/m, $V$ is the volume of the quasi-1D chain, and $E_0$ is a constant of integration.

We consider now a nanomagnet of length $l$, width $w$, and thickness $d$, split in three unequal pieces with boundaries at $l_1$ and $l_2$ so that their volumes are $V_1=wtl_1$, $V_2=wt(l_2-l_1)=wt\Delta l_{1,2}$ and $V_3=wt(l_3-l_2)=wt\Delta l_{2,3}$. This scenario is schematically shown in Fig.~\ref{fig:StadiumSplit}. The total exchange energy is
\begin{equation}
\label{eq:18}
    E_\mathrm{ex} = J_1\left(\mathbf{m}_1\cdot\mathbf{m}_2\right)+J_2\left(\mathbf{m}_2\cdot\mathbf{m}_3\right),
\end{equation}
where the magnetization vectors are taken in the geometric center of each piece. This leads to $\mathbf{m}_1\cdot\mathbf{m}_2=\cos{(k_0\Delta l_{1,2})}$ and $\mathbf{m}_2\cdot\mathbf{m}_3=\cos{(k_0\Delta l_{2,3})}$. Expanding the cosine to first order in Eq.~\eqref{eq:18} and equating to the continuum solution, we obtain
\begin{equation}
\label{eq:19}
    E_\mathrm{ex} = \left(J_1+J_2\right)+\left(J_1\frac{\Delta l^2_{1,2}}{2}+J_2\frac{\Delta l^2_{2,3}}{2}\right)k_0^2=E_0+AVk_0^2,
\end{equation}

Given that the exchange constant is uniform in the nanomagnet, we can set $J_1=C_{1,2}A$ and $J_2=C_{2,3}A$. From geometry, it can be shown that
\begin{subequations}
\begin{eqnarray}
\label{eq:20_1}
    C_{1,2} &=& \frac{2}{\Delta l^2_{1,2}}\left(V_1+\frac{V_2}{2}\right),\\
\label{eq:20_2}
    C_{2,3} &=& \frac{2}{\Delta l^2_{2,3}}\left(V_3+\frac{V_2}{2}\right).
\end{eqnarray}
\end{subequations}

In the case of stadium-shaped nanomagnets, one can consider a symmetric splitting so that $V_1=V_3=V_e$ and $V_2=V_b$, leading to the expression shown in Eq.~\eqref{eq:21}.

The edge volume $V_e$ and the bulk volume $V_b$ can be computed as a function of $\Delta l$. We have two cases.

\subsubsection{Case $l/4<\Delta l <(2l-w)/4$}

This corresponds to the situation where an edge mode occupies more than the half-circle in the stadium's edge at the expense of the bulk mode. 
The edge and bulk volumes are
\begin{subequations}
\begin{eqnarray}
\label{eq:22_C1e}
    V_e &=& \frac{1}{2}\left[wtl-w^2t\left(1-\frac{\pi}{4}\right)-V_b\right],\\
\label{eq:22_C1b}
    V_b &=& (4\Delta l-l)wt.
\end{eqnarray}
\end{subequations}

\subsubsection{Case $(2l-w)/4 \leq \Delta l <l/2$}

This corresponds to the situation where an edge mode is confined to the half-circle in the stadium's edge. 
Computing the cone angle
\begin{equation}
\label{eq:22_C2theta}
    \theta = 2\arccos{\left(1-\frac{2l-4\Delta l}{w}\right)},
\end{equation}
the edge and bulk volumes are
\begin{subequations}
\begin{eqnarray}
\label{eq:22_C2e}
    V_e &=& \frac{\theta-\sin{(\theta)}}{8}w^2t,\\
\label{eq:22_C2b}
    V_b &=& wtl-w^2t\left(1-\frac{\pi}{4}\right)-2V_e.
\end{eqnarray}
\end{subequations}

The edge and bulk volumes as a function of $\Delta l$ are shown in Fig.~\ref{fig:volumes}. The limiting case $\Delta l = (2l-w)/4$ is considered to be the default.
\begin{figure}[t]
\centering \includegraphics[width=3in]{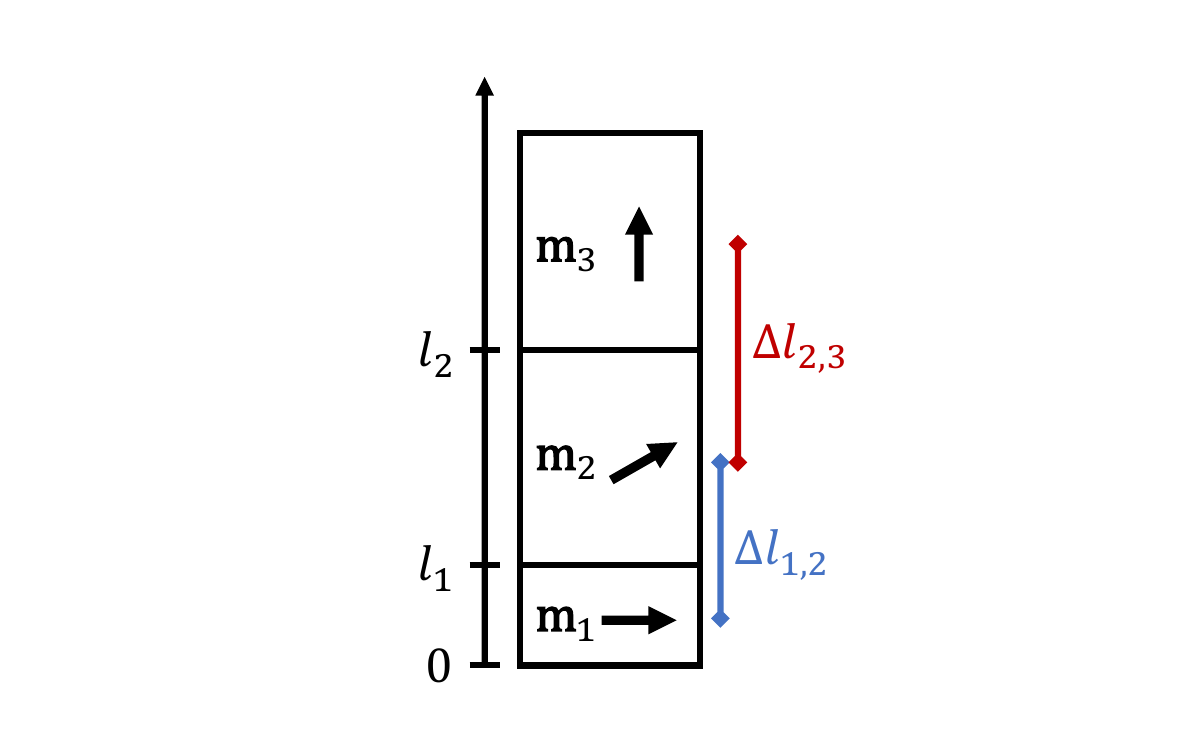}
\caption{ \label{fig:StadiumSplit} Toy model for a quasi-1D spin chain splitted into unequal pieces to estimate the exchange energy. }
\end{figure}
\begin{figure}[t]
\centering \includegraphics[width=3.3in]{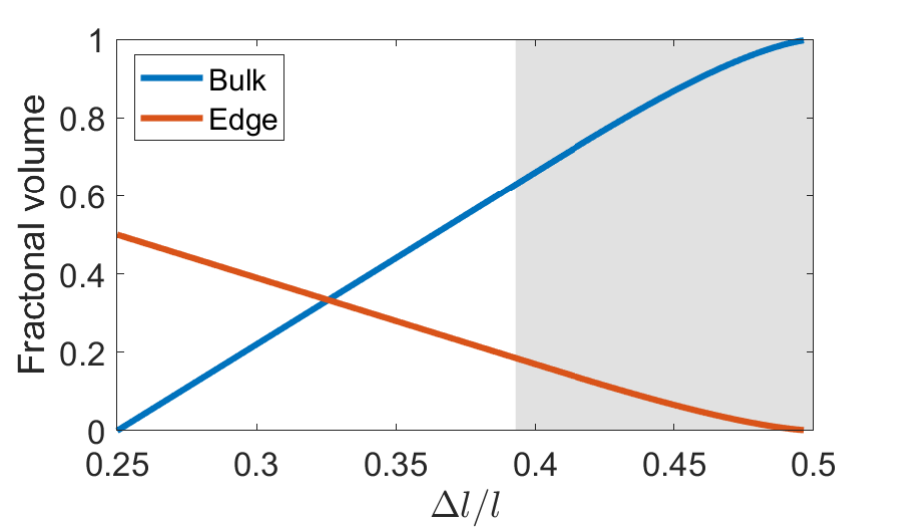}
\caption{ \label{fig:volumes} Fractional ratio between edge and volume modes. The default is considered at the edge of the gray area in the limiting case $\Delta l = (2l-w)/4$. }
\end{figure}
 \begin{figure*}[t]
 \centering \includegraphics[width=6.5in]{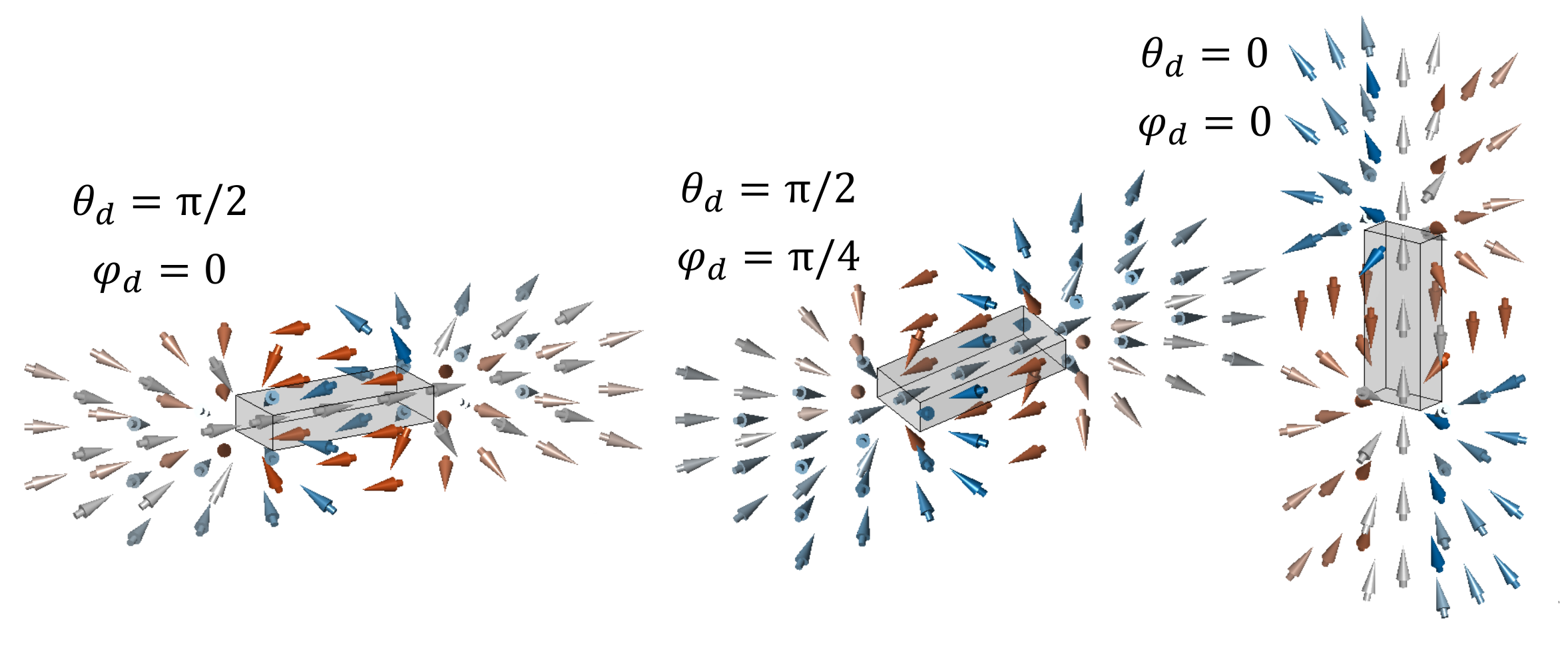}
 \caption{ \label{fig:appstray} Stray fields computed with Eqs.~\eqref{eq:stray}. We show the computed field under several rotations within G\ae{}nice and in the position of neighboring macrospins. The rotation angles are displayed for each figure. }
 \end{figure*}

\section{Derivation of exchange energy}
\label{sec:appstray}

In Ref.~\cite{Engel2005}, the authors considered a rectangular prisms with its geometric center at the origin of the Cartesian reference frame and sides $2x_b>2y_b>2z_b$. The resulting expressions for the stray field are:
 \begin{widetext}
\begin{subequations}
\label{eq:stray}
\begin{eqnarray}
\label{eq:stray_1}
    H_x(x,y,z) &=& \frac{M_s}{4\pi}\sum_{k,l,m=1}^2 (-1)^{k+l+m}\mathrm{ln}\left\{z+(-1)^mz_b+\sqrt{L(k,l,m)}\right\},\\
\label{eq:stray_2}
    H_y(x,y,z) &=& -\frac{M_s}{4\pi}\sum_{k,l,m=1}^2 (-1)^{k+l+m}\frac{\left[y+(-1)^ly_b\right]\left[x+(-1)^kx_b\right]}{|y+(-1)^ly_b||x+(-1)^kx_b|}\times\mathrm{arctan}\left\{\frac{|x+(-1)^kx_b|\left[z+(-1)^mz_b\right]}{|y+(-1)^ly_b|L(k,l,m)}\right\},\\
\label{eq:stray_3}
    H_z(x,y,z) &=& \frac{M_s}{4\pi}\sum_{k,l,m=1}^2 (-1)^{k+l+m}\mathrm{ln}\left\{x+(-1)^kx_b+\sqrt{L(k,l,m)}\right\},
\end{eqnarray}
\end{subequations}
 \end{widetext}
where 
\begin{equation}
    L(k,l,m) = \left[x+(-1)^kx_b\right]^2+\left[y+(-1)^ly_b\right]^2+\left[z+(-1)^mz_b\right]^2
\end{equation}
We note that the assumed orientation of the rectangular prism in Ref.~\cite{Engel2005} is different than that assumed in G\ae{}nice. For this reason, we rotate the expressions of Eq.~\eqref{eq:stray} such that the easy axis of the rectangular prism aligns with the $z$ axis. In Fig.~\ref{fig:appstray} we show the calculated stray field from rectangular prisms with different director vectors. In all cases, it is seen by inspection that the stray field is computed correctly.


\begin{thebibliography}{37}%
\makeatletter
\providecommand \@ifxundefined [1]{%
 \@ifx{#1\undefined}
}%
\providecommand \@ifnum [1]{%
 \ifnum #1\expandafter \@firstoftwo
 \else \expandafter \@secondoftwo
 \fi
}%
\providecommand \@ifx [1]{%
 \ifx #1\expandafter \@firstoftwo
 \else \expandafter \@secondoftwo
 \fi
}%
\providecommand \natexlab [1]{#1}%
\providecommand \enquote  [1]{``#1''}%
\providecommand \bibnamefont  [1]{#1}%
\providecommand \bibfnamefont [1]{#1}%
\providecommand \citenamefont [1]{#1}%
\providecommand \href@noop [0]{\@secondoftwo}%
\providecommand \href [0]{\begingroup \@sanitize@url \@href}%
\providecommand \@href[1]{\@@startlink{#1}\@@href}%
\providecommand \@@href[1]{\endgroup#1\@@endlink}%
\providecommand \@sanitize@url [0]{\catcode `\\12\catcode `\$12\catcode
  `\&12\catcode `\#12\catcode `\^12\catcode `\_12\catcode `\%12\relax}%
\providecommand \@@startlink[1]{}%
\providecommand \@@endlink[0]{}%
\providecommand \url  [0]{\begingroup\@sanitize@url \@url }%
\providecommand \@url [1]{\endgroup\@href {#1}{\urlprefix }}%
\providecommand \urlprefix  [0]{URL }%
\providecommand \Eprint [0]{\href }%
\providecommand \doibase [0]{http://dx.doi.org/}%
\providecommand \selectlanguage [0]{\@gobble}%
\providecommand \bibinfo  [0]{\@secondoftwo}%
\providecommand \bibfield  [0]{\@secondoftwo}%
\providecommand \translation [1]{[#1]}%
\providecommand \BibitemOpen [0]{}%
\providecommand \bibitemStop [0]{}%
\providecommand \bibitemNoStop [0]{.\EOS\space}%
\providecommand \EOS [0]{\spacefactor3000\relax}%
\providecommand \BibitemShut  [1]{\csname bibitem#1\endcsname}%
\let\auto@bib@innerbib\@empty
\bibitem [{\citenamefont {Skj{\ae}rv{\o}}\ \emph {et~al.}(2019)\citenamefont
  {Skj{\ae}rv{\o}}, \citenamefont {Marrows}, \citenamefont {Stamps},\ and\
  \citenamefont {Heyderman}}]{Skaervo2019}%
  \BibitemOpen
  \bibfield  {author} {\bibinfo {author} {\bibfnamefont {S.~H.}\ \bibnamefont
  {Skj{\ae}rv{\o}}}, \bibinfo {author} {\bibfnamefont {C.~H.}\ \bibnamefont
  {Marrows}}, \bibinfo {author} {\bibfnamefont {R.~L.}\ \bibnamefont {Stamps}},
  \ and\ \bibinfo {author} {\bibfnamefont {L.~J.}\ \bibnamefont {Heyderman}},\
  }\bibfield  {title} {\enquote {\bibinfo {title} {Advances in artificial spin
  ice},}\ }\href {\doibase 10.1038/s42254-019-0118-3} {\bibfield  {journal}
  {\bibinfo  {journal} {Nature Reviews Physics}\ } (\bibinfo {year} {2019}),\
  10.1038/s42254-019-0118-3}\BibitemShut {NoStop}%
\bibitem [{\citenamefont {Heyderman}\ and\ \citenamefont
  {Stamps}(2013)}]{Heyderman2013}%
  \BibitemOpen
  \bibfield  {author} {\bibinfo {author} {\bibfnamefont {L.~J.}\ \bibnamefont
  {Heyderman}}\ and\ \bibinfo {author} {\bibfnamefont {R.~L.}\ \bibnamefont
  {Stamps}},\ }\bibfield  {title} {\enquote {\bibinfo {title} {Artificial
  ferroic systems: Novel functionality from structure, interactions and
  dynamics},}\ }\href@noop {} {\bibfield  {journal} {\bibinfo  {journal}
  {Journal of Physics: Condensed Matter}\ }\textbf {\bibinfo {volume} {25}},\
  \bibinfo {pages} {363201} (\bibinfo {year} {2013})}\BibitemShut {NoStop}%
\bibitem [{\citenamefont {Gliga}\ \emph {et~al.}(2013)\citenamefont {Gliga},
  \citenamefont {K\'akay}, \citenamefont {Hertel},\ and\ \citenamefont
  {Heinonen}}]{Gliga2013}%
  \BibitemOpen
  \bibfield  {author} {\bibinfo {author} {\bibfnamefont {S.}~\bibnamefont
  {Gliga}}, \bibinfo {author} {\bibfnamefont {A.}~\bibnamefont {K\'akay}},
  \bibinfo {author} {\bibfnamefont {R.}~\bibnamefont {Hertel}}, \ and\ \bibinfo
  {author} {\bibfnamefont {O.~G.}\ \bibnamefont {Heinonen}},\ }\bibfield
  {title} {\enquote {\bibinfo {title} {Spectral analysis of topological defects
  in an artificial spin-ice lattice},}\ }\href {\doibase
  10.1103/PhysRevLett.110.117205} {\bibfield  {journal} {\bibinfo  {journal}
  {Phys. Rev. Lett.}\ }\textbf {\bibinfo {volume} {110}},\ \bibinfo {pages}
  {117205} (\bibinfo {year} {2013})}\BibitemShut {NoStop}%
\bibitem [{\citenamefont {Gliga}, \citenamefont {Iacocca},\ and\ \citenamefont
  {Heinonen}(2020)}]{Gliga2020}%
  \BibitemOpen
  \bibfield  {author} {\bibinfo {author} {\bibfnamefont {S.}~\bibnamefont
  {Gliga}}, \bibinfo {author} {\bibfnamefont {E.}~\bibnamefont {Iacocca}}, \
  and\ \bibinfo {author} {\bibfnamefont {O.~G.}\ \bibnamefont {Heinonen}},\
  }\bibfield  {title} {\enquote {\bibinfo {title} {Dynamics of reconfigurable
  artificial spin ice: Toward magnonic functional materials},}\ }\href
  {\doibase 10.1063/1.5142705} {\bibfield  {journal} {\bibinfo  {journal} {APL
  Materials}\ }\textbf {\bibinfo {volume} {8}},\ \bibinfo {pages} {040911}
  (\bibinfo {year} {2020})},\ \Eprint
  {http://arxiv.org/abs/https://doi.org/10.1063/1.5142705}
  {https://doi.org/10.1063/1.5142705} \BibitemShut {NoStop}%
\bibitem [{\citenamefont {Lendinez}\ and\ \citenamefont
  {Jungfleisch}(2019)}]{Lendinez2019}%
  \BibitemOpen
  \bibfield  {author} {\bibinfo {author} {\bibfnamefont {S.}~\bibnamefont
  {Lendinez}}\ and\ \bibinfo {author} {\bibfnamefont {M.~B.}\ \bibnamefont
  {Jungfleisch}},\ }\bibfield  {title} {\enquote {\bibinfo {title}
  {Magnetization dynamics in artificial spin ice},}\ }\href@noop {} {\bibfield
  {journal} {\bibinfo  {journal} {J. Phys.: Condens. Matter}\ }\textbf
  {\bibinfo {volume} {32}},\ \bibinfo {pages} {013001} (\bibinfo {year}
  {2019})}\BibitemShut {NoStop}%
\bibitem [{\citenamefont {Gartside}\ \emph {et~al.}(2018)\citenamefont
  {Gartside}, \citenamefont {Arroo}, \citenamefont {Burn}, \citenamefont
  {Bemmer}, \citenamefont {Moskalenko}, \citenamefont {Cohen},\ and\
  \citenamefont {Branford}}]{Gartside2018}%
  \BibitemOpen
  \bibfield  {author} {\bibinfo {author} {\bibfnamefont {J.~C.}\ \bibnamefont
  {Gartside}}, \bibinfo {author} {\bibfnamefont {D.~M.}\ \bibnamefont {Arroo}},
  \bibinfo {author} {\bibfnamefont {D.~M.}\ \bibnamefont {Burn}}, \bibinfo
  {author} {\bibfnamefont {V.~L.}\ \bibnamefont {Bemmer}}, \bibinfo {author}
  {\bibfnamefont {A.}~\bibnamefont {Moskalenko}}, \bibinfo {author}
  {\bibfnamefont {L.~F.}\ \bibnamefont {Cohen}}, \ and\ \bibinfo {author}
  {\bibfnamefont {W.~R.}\ \bibnamefont {Branford}},\ }\bibfield  {title}
  {\enquote {\bibinfo {title} {Realization of ground state in artificial kagome
  spin ice via topological defect-driven magnetic writing},}\ }\href@noop {}
  {\bibfield  {journal} {\bibinfo  {journal} {Nature Nanotechnology}\ }\textbf
  {\bibinfo {volume} {13}},\ \bibinfo {pages} {53?58} (\bibinfo {year}
  {2018})}\BibitemShut {NoStop}%
\bibitem [{\citenamefont {Mamica}\ \emph {et~al.}(2018)\citenamefont {Mamica},
  \citenamefont {Zhou}, \citenamefont {Adeyeye}, \citenamefont {Krawczyk},\
  and\ \citenamefont {Gubbiotti}}]{Mamica2018}%
  \BibitemOpen
  \bibfield  {author} {\bibinfo {author} {\bibfnamefont {S.}~\bibnamefont
  {Mamica}}, \bibinfo {author} {\bibfnamefont {X.}~\bibnamefont {Zhou}},
  \bibinfo {author} {\bibfnamefont {A.}~\bibnamefont {Adeyeye}}, \bibinfo
  {author} {\bibfnamefont {M.}~\bibnamefont {Krawczyk}}, \ and\ \bibinfo
  {author} {\bibfnamefont {G.}~\bibnamefont {Gubbiotti}},\ }\bibfield  {title}
  {\enquote {\bibinfo {title} {Spin-wave dynamics in artificial anti-spin-ice
  systems: Experimental and theoretical investigations},}\ }\href {\doibase
  10.1103/PhysRevB.98.054405} {\bibfield  {journal} {\bibinfo  {journal} {Phys.
  Rev. B}\ }\textbf {\bibinfo {volume} {98}},\ \bibinfo {pages} {054405}
  (\bibinfo {year} {2018})}\BibitemShut {NoStop}%
\bibitem [{\citenamefont {Arroo}, \citenamefont {Gartside},\ and\ \citenamefont
  {Branford}(2019)}]{Arroo2019}%
  \BibitemOpen
  \bibfield  {author} {\bibinfo {author} {\bibfnamefont {D.~M.}\ \bibnamefont
  {Arroo}}, \bibinfo {author} {\bibfnamefont {J.~C.}\ \bibnamefont {Gartside}},
  \ and\ \bibinfo {author} {\bibfnamefont {W.~R.}\ \bibnamefont {Branford}},\
  }\bibfield  {title} {\enquote {\bibinfo {title} {Sculpting the spin-wave
  response of artificial spin ice via microstate selection},}\ }\href {\doibase
  10.1103/PhysRevB.100.214425} {\bibfield  {journal} {\bibinfo  {journal}
  {Phys. Rev. B}\ }\textbf {\bibinfo {volume} {100}},\ \bibinfo {pages}
  {214425} (\bibinfo {year} {2019})}\BibitemShut {NoStop}%
\bibitem [{\citenamefont {Dion}\ \emph
  {et~al.}(2019{\natexlab{a}})\citenamefont {Dion}, \citenamefont {Arroo},
  \citenamefont {Yamanoi}, \citenamefont {Kimura}, \citenamefont {Gartside},
  \citenamefont {Cohen}, \citenamefont {Kurebayashi},\ and\ \citenamefont
  {Branford}}]{Dion2019}%
  \BibitemOpen
  \bibfield  {author} {\bibinfo {author} {\bibfnamefont {T.}~\bibnamefont
  {Dion}}, \bibinfo {author} {\bibfnamefont {D.~M.}\ \bibnamefont {Arroo}},
  \bibinfo {author} {\bibfnamefont {K.}~\bibnamefont {Yamanoi}}, \bibinfo
  {author} {\bibfnamefont {T.}~\bibnamefont {Kimura}}, \bibinfo {author}
  {\bibfnamefont {J.~C.}\ \bibnamefont {Gartside}}, \bibinfo {author}
  {\bibfnamefont {L.~F.}\ \bibnamefont {Cohen}}, \bibinfo {author}
  {\bibfnamefont {H.}~\bibnamefont {Kurebayashi}}, \ and\ \bibinfo {author}
  {\bibfnamefont {W.~R.}\ \bibnamefont {Branford}},\ }\bibfield  {title}
  {\enquote {\bibinfo {title} {Tunable magnetization dynamics in artificial
  spin ice via shape anisotropy modification},}\ }\href@noop {} {\bibfield
  {journal} {\bibinfo  {journal} {Phys. Rev. B}\ }\textbf {\bibinfo {volume}
  {100}},\ \bibinfo {pages} {054433} (\bibinfo {year}
  {2019}{\natexlab{a}})}\BibitemShut {NoStop}%
\bibitem [{\citenamefont {Iacocca}, \citenamefont {Gliga},\ and\ \citenamefont
  {Heinonen}(2020)}]{Iacocca2020}%
  \BibitemOpen
  \bibfield  {author} {\bibinfo {author} {\bibfnamefont {E.}~\bibnamefont
  {Iacocca}}, \bibinfo {author} {\bibfnamefont {S.}~\bibnamefont {Gliga}}, \
  and\ \bibinfo {author} {\bibfnamefont {O.~G.}\ \bibnamefont {Heinonen}},\
  }\bibfield  {title} {\enquote {\bibinfo {title} {Tailoring spin-wave channels
  in a reconfigurable artificial spin ice},}\ }\href {\doibase
  10.1103/PhysRevApplied.13.044047} {\bibfield  {journal} {\bibinfo  {journal}
  {Phys. Rev. Applied}\ }\textbf {\bibinfo {volume} {13}},\ \bibinfo {pages}
  {044047} (\bibinfo {year} {2020})}\BibitemShut {NoStop}%
\bibitem [{\citenamefont {Micaletti}\ and\ \citenamefont
  {Montoncello}(2023)}]{MIcaletti2023}%
  \BibitemOpen
  \bibfield  {author} {\bibinfo {author} {\bibfnamefont {P.}~\bibnamefont
  {Micaletti}}\ and\ \bibinfo {author} {\bibfnamefont {F.}~\bibnamefont
  {Montoncello}},\ }\bibfield  {title} {\enquote {\bibinfo {title} {Dynamic
  footprints of the specific artificial spin ice microstate on its spin
  waves},}\ }\href {\doibase 10.3390/magnetochemistry9060158} {\bibfield
  {journal} {\bibinfo  {journal} {Magnetochemistry}\ }\textbf {\bibinfo
  {volume} {9}} (\bibinfo {year} {2023}),\
  10.3390/magnetochemistry9060158}\BibitemShut {NoStop}%
\bibitem [{\citenamefont {Lendinez}\ \emph {et~al.}(2021)\citenamefont
  {Lendinez}, \citenamefont {Kaffash}, \citenamefont {Heinonen}, \citenamefont
  {Gliga}, \citenamefont {Iacocca},\ and\ \citenamefont
  {Jungfleisch}}]{Lendinez2023}%
  \BibitemOpen
  \bibfield  {author} {\bibinfo {author} {\bibfnamefont {S.}~\bibnamefont
  {Lendinez}}, \bibinfo {author} {\bibfnamefont {M.~T.}\ \bibnamefont
  {Kaffash}}, \bibinfo {author} {\bibfnamefont {O.~G.}\ \bibnamefont
  {Heinonen}}, \bibinfo {author} {\bibfnamefont {S.}~\bibnamefont {Gliga}},
  \bibinfo {author} {\bibfnamefont {E.}~\bibnamefont {Iacocca}}, \ and\
  \bibinfo {author} {\bibfnamefont {M.~B.}\ \bibnamefont {Jungfleisch}},\
  }\bibfield  {title} {\enquote {\bibinfo {title} {Nonlinear multi-magnon
  scattering in artificial spin ice},}\ }\href
  {https://www.nature.com/articles/s41467-023-38992-7} {\bibfield  {journal}
  {\bibinfo  {journal} {Nature Communications}\ }\textbf {\bibinfo {volume}
  {14}},\ \bibinfo {pages} {3419} (\bibinfo {year} {2021})}\BibitemShut
  {NoStop}%
\bibitem [{\citenamefont {Iacocca}\ \emph {et~al.}(2016)\citenamefont
  {Iacocca}, \citenamefont {Gliga}, \citenamefont {Stamps},\ and\ \citenamefont
  {Heinonen}}]{Iacocca2016}%
  \BibitemOpen
  \bibfield  {author} {\bibinfo {author} {\bibfnamefont {E.}~\bibnamefont
  {Iacocca}}, \bibinfo {author} {\bibfnamefont {S.}~\bibnamefont {Gliga}},
  \bibinfo {author} {\bibfnamefont {R.~L.}\ \bibnamefont {Stamps}}, \ and\
  \bibinfo {author} {\bibfnamefont {O.}~\bibnamefont {Heinonen}},\ }\bibfield
  {title} {\enquote {\bibinfo {title} {Reconfigurable wave band structure of an
  artificial square ice},}\ }\href {\doibase 10.1103/PhysRevB.93.134420}
  {\bibfield  {journal} {\bibinfo  {journal} {Phys. Rev. B}\ }\textbf {\bibinfo
  {volume} {93}},\ \bibinfo {pages} {134420} (\bibinfo {year}
  {2016})}\BibitemShut {NoStop}%
\bibitem [{\citenamefont {Lasnier}\ and\ \citenamefont
  {Wysin}(2020)}]{Lasnier2020}%
  \BibitemOpen
  \bibfield  {author} {\bibinfo {author} {\bibfnamefont {T.~D.}\ \bibnamefont
  {Lasnier}}\ and\ \bibinfo {author} {\bibfnamefont {G.~M.}\ \bibnamefont
  {Wysin}},\ }\bibfield  {title} {\enquote {\bibinfo {title} {Magnetic
  oscillation modes in square-lattice artificial spin ice},}\ }\href {\doibase
  10.1103/PhysRevB.101.224428} {\bibfield  {journal} {\bibinfo  {journal}
  {Phys. Rev. B}\ }\textbf {\bibinfo {volume} {101}},\ \bibinfo {pages}
  {224428} (\bibinfo {year} {2020})}\BibitemShut {NoStop}%
\bibitem [{\citenamefont {Montoncello}\ \emph {et~al.}(2023)\citenamefont
  {Montoncello}, \citenamefont {Kaffash}, \citenamefont {Carfagno},
  \citenamefont {Doty}, \citenamefont {Gubbiotti},\ and\ \citenamefont
  {Jungfleisch}}]{Montoncello2023}%
  \BibitemOpen
  \bibfield  {author} {\bibinfo {author} {\bibfnamefont {F.}~\bibnamefont
  {Montoncello}}, \bibinfo {author} {\bibfnamefont {M.~T.}\ \bibnamefont
  {Kaffash}}, \bibinfo {author} {\bibfnamefont {H.}~\bibnamefont {Carfagno}},
  \bibinfo {author} {\bibfnamefont {M.~F.}\ \bibnamefont {Doty}}, \bibinfo
  {author} {\bibfnamefont {G.}~\bibnamefont {Gubbiotti}}, \ and\ \bibinfo
  {author} {\bibfnamefont {M.~B.}\ \bibnamefont {Jungfleisch}},\ }\bibfield
  {title} {\enquote {\bibinfo {title} {{A Brillouin light scattering study of
  the spin-wave magnetic field dependence in a magnetic hybrid system made of
  an artificial spin-ice structure and a film underlayer}},}\ }\href {\doibase
  10.1063/5.0140866} {\bibfield  {journal} {\bibinfo  {journal} {Journal of
  Applied Physics}\ }\textbf {\bibinfo {volume} {133}},\ \bibinfo {pages}
  {083901} (\bibinfo {year} {2023})}\BibitemShut {NoStop}%
\bibitem [{\citenamefont {Graczyk}\ \emph {et~al.}(2018)\citenamefont
  {Graczyk}, \citenamefont {Krawczyk}, \citenamefont {Dhuey}, \citenamefont
  {Yang}, \citenamefont {Schmidt},\ and\ \citenamefont
  {Gubbiotti}}]{Graczyk2018}%
  \BibitemOpen
  \bibfield  {author} {\bibinfo {author} {\bibfnamefont {P.}~\bibnamefont
  {Graczyk}}, \bibinfo {author} {\bibfnamefont {M.}~\bibnamefont {Krawczyk}},
  \bibinfo {author} {\bibfnamefont {S.}~\bibnamefont {Dhuey}}, \bibinfo
  {author} {\bibfnamefont {W.-G.}\ \bibnamefont {Yang}}, \bibinfo {author}
  {\bibfnamefont {H.}~\bibnamefont {Schmidt}}, \ and\ \bibinfo {author}
  {\bibfnamefont {G.}~\bibnamefont {Gubbiotti}},\ }\bibfield  {title} {\enquote
  {\bibinfo {title} {Magnonic band gap and mode hybridization in continuous
  permalloy films induced by vertical dynamic coupling with an array of
  permalloy ellipses},}\ }\href {\doibase 10.1103/PhysRevB.98.174420}
  {\bibfield  {journal} {\bibinfo  {journal} {Phys. Rev. B}\ }\textbf {\bibinfo
  {volume} {98}},\ \bibinfo {pages} {174420} (\bibinfo {year}
  {2018})}\BibitemShut {NoStop}%
\bibitem [{\citenamefont {Negrello}\ \emph {et~al.}(2022)\citenamefont
  {Negrello}, \citenamefont {Montoncello}, \citenamefont {Kaffash},
  \citenamefont {Jungfleisch},\ and\ \citenamefont {Gubbiotti}}]{Negrello2022}%
  \BibitemOpen
  \bibfield  {author} {\bibinfo {author} {\bibfnamefont {R.}~\bibnamefont
  {Negrello}}, \bibinfo {author} {\bibfnamefont {F.}~\bibnamefont
  {Montoncello}}, \bibinfo {author} {\bibfnamefont {M.~T.}\ \bibnamefont
  {Kaffash}}, \bibinfo {author} {\bibfnamefont {M.~B.}\ \bibnamefont
  {Jungfleisch}}, \ and\ \bibinfo {author} {\bibfnamefont {G.}~\bibnamefont
  {Gubbiotti}},\ }\bibfield  {title} {\enquote {\bibinfo {title} {Dynamic
  coupling and spin-wave dispersions in a magnetic hybrid system made of an
  artificial spin-ice structure and an extended nife underlayer},}\ }\href
  {\doibase 10.1063/5.0102571} {\bibfield  {journal} {\bibinfo  {journal} {APL
  Materials}\ }\textbf {\bibinfo {volume} {10}},\ \bibinfo {pages} {091115}
  (\bibinfo {year} {2022})}\BibitemShut {NoStop}%
\bibitem [{\citenamefont {Wang}\ \emph {et~al.}(2006)\citenamefont {Wang},
  \citenamefont {Nisoli}, \citenamefont {Freitas}, \citenamefont {Li},
  \citenamefont {McConville}, \citenamefont {Cooley}, \citenamefont {Lund},
  \citenamefont {Samarth}, \citenamefont {Leighton}, \citenamefont {Crespi},\
  and\ \citenamefont {Schiffer}}]{Wang2006}%
  \BibitemOpen
  \bibfield  {author} {\bibinfo {author} {\bibfnamefont {R.~F.}\ \bibnamefont
  {Wang}}, \bibinfo {author} {\bibfnamefont {C.}~\bibnamefont {Nisoli}},
  \bibinfo {author} {\bibfnamefont {R.~S.}\ \bibnamefont {Freitas}}, \bibinfo
  {author} {\bibfnamefont {J.}~\bibnamefont {Li}}, \bibinfo {author}
  {\bibfnamefont {W.}~\bibnamefont {McConville}}, \bibinfo {author}
  {\bibfnamefont {B.~J.}\ \bibnamefont {Cooley}}, \bibinfo {author}
  {\bibfnamefont {M.~S.}\ \bibnamefont {Lund}}, \bibinfo {author}
  {\bibfnamefont {N.}~\bibnamefont {Samarth}}, \bibinfo {author} {\bibfnamefont
  {C.}~\bibnamefont {Leighton}}, \bibinfo {author} {\bibfnamefont {V.~H.}\
  \bibnamefont {Crespi}}, \ and\ \bibinfo {author} {\bibfnamefont
  {P.}~\bibnamefont {Schiffer}},\ }\bibfield  {title} {\enquote {\bibinfo
  {title} {Artificial spin ice in a geometrically frustrated lattice of
  nanoscale ferromagnetic islands},}\ }\href@noop {} {\bibfield  {journal}
  {\bibinfo  {journal} {Nature}\ }\textbf {\bibinfo {volume} {439}},\ \bibinfo
  {pages} {303--306} (\bibinfo {year} {2006})}\BibitemShut {NoStop}%
\bibitem [{\citenamefont {Iacocca}\ and\ \citenamefont
  {Heinonen}(2017)}]{Iacocca2017c}%
  \BibitemOpen
  \bibfield  {author} {\bibinfo {author} {\bibfnamefont {E.}~\bibnamefont
  {Iacocca}}\ and\ \bibinfo {author} {\bibfnamefont {O.}~\bibnamefont
  {Heinonen}},\ }\bibfield  {title} {\enquote {\bibinfo {title} {Topologically
  nontrivial magnon bands in artificial square spin ices with
  dzyaloshinskii-moriya interaction},}\ }\href@noop {} {\bibfield  {journal}
  {\bibinfo  {journal} {Phys. Rev. Applied}\ }\textbf {\bibinfo {volume} {8}},\
  \bibinfo {pages} {034015} (\bibinfo {year} {2017})}\BibitemShut {NoStop}%
\bibitem [{\citenamefont {Dion}\ \emph
  {et~al.}(2019{\natexlab{b}})\citenamefont {Dion}, \citenamefont {Stenning},
  \citenamefont {Vanstone}, \citenamefont {Holder}, \citenamefont {Sultana},
  \citenamefont {Alatteili}, \citenamefont {Martinez}, \citenamefont {Kaffash},
  \citenamefont {Kimura}, \citenamefont {Kurebayashi}, \citenamefont
  {Branford}, \citenamefont {Iacocca}, \citenamefont {Jungfleisch},\ and\
  \citenamefont {Gartside}}]{Dion2023}%
  \BibitemOpen
  \bibfield  {author} {\bibinfo {author} {\bibfnamefont {T.}~\bibnamefont
  {Dion}}, \bibinfo {author} {\bibfnamefont {K.~D.}\ \bibnamefont {Stenning}},
  \bibinfo {author} {\bibfnamefont {A.}~\bibnamefont {Vanstone}}, \bibinfo
  {author} {\bibfnamefont {H.~H.}\ \bibnamefont {Holder}}, \bibinfo {author}
  {\bibfnamefont {R.}~\bibnamefont {Sultana}}, \bibinfo {author} {\bibfnamefont
  {G.}~\bibnamefont {Alatteili}}, \bibinfo {author} {\bibfnamefont
  {V.}~\bibnamefont {Martinez}}, \bibinfo {author} {\bibfnamefont {M.~T.}\
  \bibnamefont {Kaffash}}, \bibinfo {author} {\bibfnamefont {T.}~\bibnamefont
  {Kimura}}, \bibinfo {author} {\bibfnamefont {H.}~\bibnamefont {Kurebayashi}},
  \bibinfo {author} {\bibfnamefont {W.~R.}\ \bibnamefont {Branford}}, \bibinfo
  {author} {\bibfnamefont {E.}~\bibnamefont {Iacocca}}, \bibinfo {author}
  {\bibfnamefont {B.~M.}\ \bibnamefont {Jungfleisch}}, \ and\ \bibinfo {author}
  {\bibfnamefont {J.~C.}\ \bibnamefont {Gartside}},\ }\bibfield  {title}
  {\enquote {\bibinfo {title} {Ultrastrong magnon-magnon coupling and chiral
  symmetry breaking in a 3d magnonic metamaterial},}\ }\href@noop {} {\bibfield
   {journal} {\bibinfo  {journal} {arXiv:2306.16159}\ } (\bibinfo {year}
  {2019}{\natexlab{b}})}\BibitemShut {NoStop}%
\bibitem [{\citenamefont {Abert}(2019)}]{Abert2019}%
  \BibitemOpen
  \bibfield  {author} {\bibinfo {author} {\bibfnamefont {C.}~\bibnamefont
  {Abert}},\ }\bibfield  {title} {\enquote {\bibinfo {title} {Micromagnetics
  and spintronics: models and numerical methods},}\ }\href@noop {} {\bibfield
  {journal} {\bibinfo  {journal} {The European Physical Journal B}\ }\textbf
  {\bibinfo {volume} {92}},\ \bibinfo {pages} {120} (\bibinfo {year}
  {2019})}\BibitemShut {NoStop}%
\bibitem [{\citenamefont {May}\ \emph {et~al.}(2019)\citenamefont {May},
  \citenamefont {Hunt}, \citenamefont {Van Den~Berg}, \citenamefont {Hejazi},\
  and\ \citenamefont {Ladak}}]{May2019}%
  \BibitemOpen
  \bibfield  {author} {\bibinfo {author} {\bibfnamefont {A.}~\bibnamefont
  {May}}, \bibinfo {author} {\bibfnamefont {M.}~\bibnamefont {Hunt}}, \bibinfo
  {author} {\bibfnamefont {A.}~\bibnamefont {Van Den~Berg}}, \bibinfo {author}
  {\bibfnamefont {A.}~\bibnamefont {Hejazi}}, \ and\ \bibinfo {author}
  {\bibfnamefont {S.}~\bibnamefont {Ladak}},\ }\bibfield  {title} {\enquote
  {\bibinfo {title} {Realisation of a frustrated 3d magnetic nanowire
  lattice},}\ }\href {https://www.nature.com/articles/s42005-018-0104-6}
  {\bibfield  {journal} {\bibinfo  {journal} {Communication Physics}\ }\textbf
  {\bibinfo {volume} {2}},\ \bibinfo {pages} {13} (\bibinfo {year}
  {2019})}\BibitemShut {NoStop}%
\bibitem [{\citenamefont {May}\ \emph {et~al.}(2021)\citenamefont {May},
  \citenamefont {Saccone}, \citenamefont {van~den Berg}, \citenamefont {Askey},
  \citenamefont {Hunt},\ and\ \citenamefont {Ladak}}]{May2021}%
  \BibitemOpen
  \bibfield  {author} {\bibinfo {author} {\bibfnamefont {A.}~\bibnamefont
  {May}}, \bibinfo {author} {\bibfnamefont {M.}~\bibnamefont {Saccone}},
  \bibinfo {author} {\bibfnamefont {A.}~\bibnamefont {van~den Berg}}, \bibinfo
  {author} {\bibfnamefont {J.}~\bibnamefont {Askey}}, \bibinfo {author}
  {\bibfnamefont {M.}~\bibnamefont {Hunt}}, \ and\ \bibinfo {author}
  {\bibfnamefont {S.}~\bibnamefont {Ladak}},\ }\bibfield  {title} {\enquote
  {\bibinfo {title} {Magnetic charge propagation upon a 3d artificial spin
  ice},}\ }\href {\doibase 10.1038/s41467-021-23480-7} {\bibfield  {journal}
  {\bibinfo  {journal} {Nature Communications}\ }\textbf {\bibinfo {volume}
  {12}},\ \bibinfo {pages} {3217} (\bibinfo {year} {2021})}\BibitemShut
  {NoStop}%
\bibitem [{\citenamefont {Sahoo}\ \emph {et~al.}(2021)\citenamefont {Sahoo},
  \citenamefont {May}, \citenamefont {van Den~Berg}, \citenamefont {Mondal},
  \citenamefont {Ladak},\ and\ \citenamefont {Barman}}]{Sahoo2021}%
  \BibitemOpen
  \bibfield  {author} {\bibinfo {author} {\bibfnamefont {S.}~\bibnamefont
  {Sahoo}}, \bibinfo {author} {\bibfnamefont {A.}~\bibnamefont {May}}, \bibinfo
  {author} {\bibfnamefont {A.}~\bibnamefont {van Den~Berg}}, \bibinfo {author}
  {\bibfnamefont {A.~K.}\ \bibnamefont {Mondal}}, \bibinfo {author}
  {\bibfnamefont {S.}~\bibnamefont {Ladak}}, \ and\ \bibinfo {author}
  {\bibfnamefont {A.}~\bibnamefont {Barman}},\ }\bibfield  {title} {\enquote
  {\bibinfo {title} {Observation of coherent spin waves in a three-dimensional
  artificial spin ice structure},}\ }\href {\doibase
  10.1021/acs.nanolett.1c00650} {\bibfield  {journal} {\bibinfo  {journal}
  {Nano Letters}\ }\textbf {\bibinfo {volume} {21}},\ \bibinfo {pages}
  {4629--4635} (\bibinfo {year} {2021})}\BibitemShut {NoStop}%
\bibitem [{\citenamefont {Slavin}\ and\ \citenamefont
  {Tiberkevich}(2009)}]{Slavin2009}%
  \BibitemOpen
  \bibfield  {author} {\bibinfo {author} {\bibfnamefont {A.}~\bibnamefont
  {Slavin}}\ and\ \bibinfo {author} {\bibfnamefont {V.}~\bibnamefont
  {Tiberkevich}},\ }\bibfield  {title} {\enquote {\bibinfo {title} {Nonlinear
  auto-oscillator theory of microwave generation by spin-polarized current},}\
  }\href@noop {} {\bibfield  {journal} {\bibinfo  {journal} {Magnetics, IEEE
  Transactions on}\ }\textbf {\bibinfo {volume} {45}},\ \bibinfo {pages} {1875
  --1918} (\bibinfo {year} {2009})}\BibitemShut {NoStop}%
\bibitem [{\citenamefont {Grimsditch}\ \emph {et~al.}(2004)\citenamefont
  {Grimsditch}, \citenamefont {Giovannini}, \citenamefont {Montoncello},
  \citenamefont {Nizzoli}, \citenamefont {Leaf},\ and\ \citenamefont
  {Kaper}}]{Grimsditch2004}%
  \BibitemOpen
  \bibfield  {author} {\bibinfo {author} {\bibfnamefont {M.}~\bibnamefont
  {Grimsditch}}, \bibinfo {author} {\bibfnamefont {L.}~\bibnamefont
  {Giovannini}}, \bibinfo {author} {\bibfnamefont {F.}~\bibnamefont
  {Montoncello}}, \bibinfo {author} {\bibfnamefont {F.}~\bibnamefont
  {Nizzoli}}, \bibinfo {author} {\bibfnamefont {G.~K.}\ \bibnamefont {Leaf}}, \
  and\ \bibinfo {author} {\bibfnamefont {H.~G.}\ \bibnamefont {Kaper}},\
  }\bibfield  {title} {\enquote {\bibinfo {title} {Magnetic normal modes in
  ferromagnetic nanoparticles: A dynamical matrix approach},}\ }\href@noop {}
  {\bibfield  {journal} {\bibinfo  {journal} {Phys. Rev. B}\ }\textbf {\bibinfo
  {volume} {70}},\ \bibinfo {pages} {054409} (\bibinfo {year}
  {2004})}\BibitemShut {NoStop}%
\bibitem [{\citenamefont {Neusser}\ \emph {et~al.}(2011)\citenamefont
  {Neusser}, \citenamefont {Duerr}, \citenamefont {Tacchi}, \citenamefont
  {Madami}, \citenamefont {Sokolovskyy}, \citenamefont {Gubbiotti},
  \citenamefont {Krawczyk},\ and\ \citenamefont {Grundler}}]{Neusser2011}%
  \BibitemOpen
  \bibfield  {author} {\bibinfo {author} {\bibfnamefont {S.}~\bibnamefont
  {Neusser}}, \bibinfo {author} {\bibfnamefont {G.}~\bibnamefont {Duerr}},
  \bibinfo {author} {\bibfnamefont {S.}~\bibnamefont {Tacchi}}, \bibinfo
  {author} {\bibfnamefont {M.}~\bibnamefont {Madami}}, \bibinfo {author}
  {\bibfnamefont {M.~L.}\ \bibnamefont {Sokolovskyy}}, \bibinfo {author}
  {\bibfnamefont {G.}~\bibnamefont {Gubbiotti}}, \bibinfo {author}
  {\bibfnamefont {M.}~\bibnamefont {Krawczyk}}, \ and\ \bibinfo {author}
  {\bibfnamefont {D.}~\bibnamefont {Grundler}},\ }\bibfield  {title} {\enquote
  {\bibinfo {title} {Magnonic minibands in antidot lattices with large
  spin-wave propagation velocities},}\ }\href@noop {} {\bibfield  {journal}
  {\bibinfo  {journal} {Phys. Rev. B}\ }\textbf {\bibinfo {volume} {84}},\
  \bibinfo {pages} {094454} (\bibinfo {year} {2011})}\BibitemShut {NoStop}%
\bibitem [{\citenamefont {Rych\l{}y}\ \emph {et~al.}(2015)\citenamefont
  {Rych\l{}y}, \citenamefont {K\l{}os}, \citenamefont {Mruczkiewicz},\ and\
  \citenamefont {Krawczyk}}]{Rychly2015}%
  \BibitemOpen
  \bibfield  {author} {\bibinfo {author} {\bibfnamefont {J.}~\bibnamefont
  {Rych\l{}y}}, \bibinfo {author} {\bibfnamefont {J.~W.}\ \bibnamefont
  {K\l{}os}}, \bibinfo {author} {\bibfnamefont {M.}~\bibnamefont
  {Mruczkiewicz}}, \ and\ \bibinfo {author} {\bibfnamefont {M.}~\bibnamefont
  {Krawczyk}},\ }\bibfield  {title} {\enquote {\bibinfo {title} {Spin waves in
  one-dimensional bicomponent magnonic quasicrystals},}\ }\href@noop {}
  {\bibfield  {journal} {\bibinfo  {journal} {Phys. Rev. B}\ }\textbf {\bibinfo
  {volume} {92}},\ \bibinfo {pages} {054414} (\bibinfo {year}
  {2015})}\BibitemShut {NoStop}%
\bibitem [{\citenamefont {Gubbiotti}\ \emph {et~al.}(2018)\citenamefont
  {Gubbiotti}, \citenamefont {Zhou}, \citenamefont {Haghshenasfard},
  \citenamefont {Cottam},\ and\ \citenamefont {Adeyeye}}]{Gubbiotti2018}%
  \BibitemOpen
  \bibfield  {author} {\bibinfo {author} {\bibfnamefont {G.}~\bibnamefont
  {Gubbiotti}}, \bibinfo {author} {\bibfnamefont {X.}~\bibnamefont {Zhou}},
  \bibinfo {author} {\bibfnamefont {Z.}~\bibnamefont {Haghshenasfard}},
  \bibinfo {author} {\bibfnamefont {M.~G.}\ \bibnamefont {Cottam}}, \ and\
  \bibinfo {author} {\bibfnamefont {A.~O.}\ \bibnamefont {Adeyeye}},\
  }\bibfield  {title} {\enquote {\bibinfo {title} {Reprogrammable magnonic band
  structure of layered permalloy/cu/permalloy nanowires},}\ }\href@noop {}
  {\bibfield  {journal} {\bibinfo  {journal} {Phys. Rev. B}\ }\textbf {\bibinfo
  {volume} {97}},\ \bibinfo {pages} {134428} (\bibinfo {year}
  {2018})}\BibitemShut {NoStop}%
\bibitem [{\citenamefont {Lisiecki}\ \emph {et~al.}(2019)\citenamefont
  {Lisiecki}, \citenamefont {Rych\l{}y}, \citenamefont
  {Ku\ifmmode~\acute{s}\else \'{s}\fi{}wik}, \citenamefont
  {G\l{}owi\ifmmode~\acute{n}\else \'{n}\fi{}ski}, \citenamefont {K\l{}os},
  \citenamefont {Gro\ss{}}, \citenamefont {Tr\"ager}, \citenamefont {Bykova},
  \citenamefont {Weigand}, \citenamefont {Zelent}, \citenamefont {Goering},
  \citenamefont {Sch\"utz}, \citenamefont {Krawczyk}, \citenamefont
  {Stobiecki}, \citenamefont {Dubowik},\ and\ \citenamefont
  {Gr\"afe}}]{Lisiecki2019}%
  \BibitemOpen
  \bibfield  {author} {\bibinfo {author} {\bibfnamefont {F.}~\bibnamefont
  {Lisiecki}}, \bibinfo {author} {\bibfnamefont {J.}~\bibnamefont {Rych\l{}y}},
  \bibinfo {author} {\bibfnamefont {P.}~\bibnamefont {Ku\ifmmode~\acute{s}\else
  \'{s}\fi{}wik}}, \bibinfo {author} {\bibfnamefont {H.}~\bibnamefont
  {G\l{}owi\ifmmode~\acute{n}\else \'{n}\fi{}ski}}, \bibinfo {author}
  {\bibfnamefont {J.~W.}\ \bibnamefont {K\l{}os}}, \bibinfo {author}
  {\bibfnamefont {F.}~\bibnamefont {Gro\ss{}}}, \bibinfo {author}
  {\bibfnamefont {N.}~\bibnamefont {Tr\"ager}}, \bibinfo {author}
  {\bibfnamefont {I.}~\bibnamefont {Bykova}}, \bibinfo {author} {\bibfnamefont
  {M.}~\bibnamefont {Weigand}}, \bibinfo {author} {\bibfnamefont
  {M.}~\bibnamefont {Zelent}}, \bibinfo {author} {\bibfnamefont {E.~J.}\
  \bibnamefont {Goering}}, \bibinfo {author} {\bibfnamefont {G.}~\bibnamefont
  {Sch\"utz}}, \bibinfo {author} {\bibfnamefont {M.}~\bibnamefont {Krawczyk}},
  \bibinfo {author} {\bibfnamefont {F.}~\bibnamefont {Stobiecki}}, \bibinfo
  {author} {\bibfnamefont {J.}~\bibnamefont {Dubowik}}, \ and\ \bibinfo
  {author} {\bibfnamefont {J.}~\bibnamefont {Gr\"afe}},\ }\bibfield  {title}
  {\enquote {\bibinfo {title} {Magnons in a quasicrystal: Propagation,
  extinction, and localization of spin waves in fibonacci structures},}\
  }\href@noop {} {\bibfield  {journal} {\bibinfo  {journal} {Phys. Rev.
  Applied}\ }\textbf {\bibinfo {volume} {11}},\ \bibinfo {pages} {054061}
  (\bibinfo {year} {2019})}\BibitemShut {NoStop}%
\bibitem [{\citenamefont {Gliga}\ \emph {et~al.}(2015)\citenamefont {Gliga},
  \citenamefont {K\'akay}, \citenamefont {Heyderman}, \citenamefont {Hertel},\
  and\ \citenamefont {Heinonen}}]{Gliga2015}%
  \BibitemOpen
  \bibfield  {author} {\bibinfo {author} {\bibfnamefont {S.}~\bibnamefont
  {Gliga}}, \bibinfo {author} {\bibfnamefont {A.}~\bibnamefont {K\'akay}},
  \bibinfo {author} {\bibfnamefont {L.~J.}\ \bibnamefont {Heyderman}}, \bibinfo
  {author} {\bibfnamefont {R.}~\bibnamefont {Hertel}}, \ and\ \bibinfo {author}
  {\bibfnamefont {O.~G.}\ \bibnamefont {Heinonen}},\ }\bibfield  {title}
  {\enquote {\bibinfo {title} {Broken vertex symmetry and finite zero-point
  entropy in the artificial square ice ground state},}\ }\href@noop {}
  {\bibfield  {journal} {\bibinfo  {journal} {Phys. Rev. B}\ }\textbf {\bibinfo
  {volume} {92}},\ \bibinfo {pages} {060413} (\bibinfo {year}
  {2015})}\BibitemShut {NoStop}%
\bibitem [{\citenamefont {Colpa}(1978)}]{Colpa1978}%
  \BibitemOpen
  \bibfield  {author} {\bibinfo {author} {\bibfnamefont {J.~H.~P.}\
  \bibnamefont {Colpa}},\ }\bibfield  {title} {\enquote {\bibinfo {title}
  {Diagonalization of the quadratic boson hamiltonian},}\ }\href@noop {}
  {\bibfield  {journal} {\bibinfo  {journal} {Physica A: Statistical Mechanics
  and its Applications}\ }\textbf {\bibinfo {volume} {93}},\ \bibinfo {pages}
  {327 -- 353} (\bibinfo {year} {1978})}\BibitemShut {NoStop}%
\bibitem [{\citenamefont {Osborn}(1945)}]{Osborn1945}%
  \BibitemOpen
  \bibfield  {author} {\bibinfo {author} {\bibfnamefont {J.~A.}\ \bibnamefont
  {Osborn}},\ }\bibfield  {title} {\enquote {\bibinfo {title} {Demagnetizing
  factors of the general ellipsoid},}\ }\href@noop {} {\bibfield  {journal}
  {\bibinfo  {journal} {Phys. Rev.}\ }\textbf {\bibinfo {volume} {67}},\
  \bibinfo {pages} {351--357} (\bibinfo {year} {1945})}\BibitemShut {NoStop}%
\bibitem [{\citenamefont {Aharoni}(1998)}]{Aharoni1998}%
  \BibitemOpen
  \bibfield  {author} {\bibinfo {author} {\bibfnamefont {A.}~\bibnamefont
  {Aharoni}},\ }\bibfield  {title} {\enquote {\bibinfo {title} {Demagnetizing
  factors for rectangular ferromagnetic prisms},}\ }\href {\doibase
  10.1063/1.367113} {\bibfield  {journal} {\bibinfo  {journal} {Journal of
  Applied Physics}\ }\textbf {\bibinfo {volume} {83}},\ \bibinfo {pages}
  {3432--3434} (\bibinfo {year} {1998})}\BibitemShut {NoStop}%
\bibitem [{\citenamefont {Martinez}\ and\ \citenamefont
  {Iacocca}(2023)}]{Martinez2023}%
  \BibitemOpen
  \bibfield  {author} {\bibinfo {author} {\bibfnamefont {V.}~\bibnamefont
  {Martinez}}\ and\ \bibinfo {author} {\bibfnamefont {E.}~\bibnamefont
  {Iacocca}},\ }\bibfield  {title} {\enquote {\bibinfo {title} {A numerical
  method to determine demagnetization factors of stadium-shaped nanoislands},}\
  }\href@noop {} {\bibfield  {journal} {\bibinfo  {journal} {IEEE Magnetics
  Letters}\ }\textbf {\bibinfo {volume} {xx}},\ \bibinfo {pages} {x--x}
  (\bibinfo {year} {2023})}\BibitemShut {NoStop}%
\bibitem [{\citenamefont {Engel-Herbert}\ and\ \citenamefont
  {Hesjedal}(2005)}]{Engel2005}%
  \BibitemOpen
  \bibfield  {author} {\bibinfo {author} {\bibfnamefont {R.}~\bibnamefont
  {Engel-Herbert}}\ and\ \bibinfo {author} {\bibfnamefont {T.}~\bibnamefont
  {Hesjedal}},\ }\bibfield  {title} {\enquote {\bibinfo {title} {Calculation of
  the magnetic stray field of a uniaxial magnetic domain},}\ }\href {\doibase
  10.1063/1.1883308} {\bibfield  {journal} {\bibinfo  {journal} {Journal of
  Applied Physics}\ }\textbf {\bibinfo {volume} {97}},\ \bibinfo {pages}
  {074504} (\bibinfo {year} {2005})}\BibitemShut {NoStop}%
\bibitem [{\citenamefont {Saccone}\ \emph {et~al.}(2023)\citenamefont
  {Saccone}, \citenamefont {Carter-Gartside}, \citenamefont {Stenning},
  \citenamefont {Branford},\ and\ \citenamefont {Caravelli}}]{Saccone2023}%
  \BibitemOpen
  \bibfield  {author} {\bibinfo {author} {\bibfnamefont {M.}~\bibnamefont
  {Saccone}}, \bibinfo {author} {\bibfnamefont {J.}~\bibnamefont
  {Carter-Gartside}}, \bibinfo {author} {\bibfnamefont {K.}~\bibnamefont
  {Stenning}}, \bibinfo {author} {\bibfnamefont {W.~R.}\ \bibnamefont
  {Branford}}, \ and\ \bibinfo {author} {\bibfnamefont {F.}~\bibnamefont
  {Caravelli}},\ }\bibfield  {title} {\enquote {\bibinfo {title} {From vertices
  to vortices in magnetic nanoislands},}\ }\href {\doibase 10.1063/5.0131158}
  {\bibfield  {journal} {\bibinfo  {journal} {Physics of Fluids}\ }\textbf
  {\bibinfo {volume} {35}},\ \bibinfo {pages} {017101} (\bibinfo {year}
  {2023})}\BibitemShut {NoStop}%
\end{thebibliography}
%

\end{document}